\documentclass[twocolumn, trackchanges]{aastex701}

\usepackage{graphicx}
\usepackage{epstopdf}
\usepackage{subfigure}
\usepackage{color}
\usepackage{soul}
\usepackage{tabularx}
\usepackage{booktabs}
\usepackage{amsmath}
\usepackage{appendix}
\let\tablenum\relax
\usepackage{siunitx}
\usepackage{enumitem}
\usepackage{CJK}
\usepackage{hyperref} 

\newcommand{\um}{$\mu$m} 

\def\msun{\hbox{${\rm M}_{\odot}$}}
\def\mstar{\hbox{${M}_{\star}$}}
\newcommand{\logM}{$\log (M_{\star}/M_{\odot})$}
\newcommand{\zphot}{$z_\textrm{phot}$}
\newcommand{\zspec}{$z_\textrm{spec}$}
\newcommand{\UVJ}{\textit{UVJ}}

\newcommand{\OII}{\hbox{{\rm [O}\kern 0.1em{\sc ii}{\rm ]$\lambda\lambda3726,3729$}}}
\newcommand{\OIIshort}{\hbox{{\rm [O}\kern 0.1em{\sc ii}]}}
\newcommand{\OIII}{\hbox{{\rm [O}\kern 0.1em{\sc iii}{\rm ]$\lambda\lambda4959,5007$}}}
\newcommand{\OIIIshort}{\hbox{{\rm [O}\kern 0.1em{\sc iii}]}}

\newcommand{\Hbeta}{$\rm{H}\beta$}

\begin{document}

\title{MAGAZ3NE: Dust Deficiency in Ultramassive Quiescent Galaxies at $3< z<4$ with ALMA Observations}
 


\author[0000-0003-2144-2943]{Wenjun Chang}
\affiliation{Department of Physics and Astronomy, University of California, Riverside, 900 University Avenue, Riverside, CA 92521, USA}
\email[show]{wenjun.chang@email.ucr.edu}

\author[0000-0002-6572-7089]{Gillian Wilson}
\affiliation{Department of Physics, University of California, Merced, 5200 Lake Road, Merced, CA 95343, USA}
\affiliation{Department of Physics and Astronomy, University of California, Riverside, 900 University Avenue, Riverside, CA 92521, USA}
\email{gwilson@ucmerced.edu}

\author[0000-0001-6003-0541]{Ben Forrest}
\affiliation{Department of Physics and Astronomy, University of California, Davis, One Shields Avenue, Davis, CA 95616, USA}
\email{bforrest@ucdavis.edu}
 
\author[0000-0002-2446-8770]{Ian McConachie}
\affiliation{Department of Astronomy, University of Wisconsin-Madison, 475 N. Charter St., Madison, WI 53706 USA}
\email{ian.mcconachie@wisc.edu} 

\author[0000-0003-1832-4137]{Allison Noble}
\affiliation{School of Earth and Space Exploration, Arizona State University, Tempe, AZ 85287, USA}
\email{anoble5@asu.edu}

\author[0000-0002-9330-9108]{Adam Muzzin}
\affiliation{Department of Physics and Astronomy, York University, 4700, Keele Street, Toronto, ON MJ3 1P3, Canada}
\email{muzzin@yorku.ca}

\author[0000-0001-9002-3502]{Danilo Marchesini}
\affiliation{Department of Physics \& Astronomy, Tufts University, MA 02155, USA}
\email{Danilo.Marchesini@tufts.edu}

\author[0000-0003-1371-6019]{M. C. Cooper}
\affiliation{Department of Physics and Astronomy, University of California, Irvine, 4129 Frederick Reines Hall, Irvine, CA 92697, USA}
\email{cooper@uci.edu}

\author{Tracy Webb}
\affiliation{Department of Physics, McGill Space Institute, McGill University, 3600 rue University, Montr\'{e}al, Qu\'{e}bec H3A 2T8, Canada}
\email{webb@physics.mcgill.ca}

\author[0000-0003-4693-6157]{Gabriela Canalizo}
\affiliation{Department of Physics and Astronomy, University of California, Riverside, 900 University Avenue, Riverside, CA 92521, USA}
\email{gabyc@ucr.edu}

\author{Percy Gomez}
\affiliation{W.M. Keck Observatory, 65-1120 Mamalahoa Hwy., Kamuela, HI 96743, USA}
\email{pgomez@keck.hawaii.edu}

\author[0000-0003-3307-7525]{Yongda Zhu}
\affiliation{Steward Observatory, University of Arizona, 933 North Cherry Avenue, Tucson, AZ 85721, USA}
\email{yongdaz@arizona.edu}

\author{Adit Harin Edward}
\affiliation{Department of Physics and Astronomy, York University, 4700, Keele Street, Toronto, ON MJ3 1P3, Canada}
\email{ahedward@yorku.ca}

\author[0000-0002-9158-6996]{Han Lei}
\affiliation{Department of Physics, McGill Space Institute, McGill University, 3600 rue University, Montr\'{e}al, Qu\'{e}bec H3A 2T8, Canada}
\email{han.lei@mail.mcgill.ca}

\author[0000-0002-9466-2763]{Aurelien Henry}
\affiliation{Department of Physics, University of California, Merced, 5200 Lake Road, Merced, CA 95343,  USA}  
\email{ahenry13@ucmerced.edu}

\author[0000-0001-8169-7249]{Stephanie M. Urbano Stawinski}
\affiliation{Department of Physics, University of California, Santa Barbara, Santa Barbara, CA 93106, USA}
\email{sstawins@ucsb.edu}

\author[0000-0002-6505-9981]{M.E. Wisz}
\affiliation{Department of Physics, University of California, Merced, 5200 Lake Road, Merced, CA 95343, USA} 
\email{mwisz@ucmerced.edu}

 
\begin{abstract}

A major challenge in identifying massive quiescent galaxies at $z>3$ is distinguishing truly passive systems from dust-obscured star-forming galaxies, as both populations exhibit similar red ultraviolet (UV)-to-near-infrared (NIR) colors.
In this work, we present ALMA Band 7 dust-continuum observations of five ultramassive galaxies (UMGs; $\log (M_\star / M_\odot) > 11$) spectroscopically confirmed at $z_{\rm spec} > 3$ from the MAGAZ3NE survey.
Our results reveal that only one galaxy shows a faint 870 \um\ dust continuum detection, while the remaining four UMGs are undetected down to the $3\sigma$ depth . 
By incorporating ALMA constraints into the spectral energy distribution analysis, we confirm that these UV-NIR-selected systems are truly quiescent UMGs, lying more than one dex below the star-forming main sequence with \mbox{$\mathrm{\log (sSFR/Gyr^{-1}) < -1}$}, thereby ruling out the possibility of obscured star formation. 
We then estimate dust masses using both spectral energy distribution modeling and modified blackbody fitting, with consistent results between the two methods. 
We find that three UMGs have evolved into extremely dust-poor quiescent galaxies, with $M_{\mathrm{dust}}/M_\star \lesssim 10^{-4}$, while the ALMA-detected galaxy has a comparatively higher dust reservoir with \mbox{$M_{\mathrm{dust}}/M_\star \sim 10^{-3}$}.
Our results present the most massive and extremely dust-poor spectroscopically confirmed quiescent galaxies known at $3 < z < 4$, providing valuable observational constraints on rapid dust removal and quenching processes in the early universe.
Future molecular line observations will be essential to directly measure the gas content and verify the efficiency of the depletion process.

\end{abstract}



\section{Introduction} 
\label{sec:intro}


 
Over the last decade, deep optical and near-infrared surveys have revealed a growing population of massive galaxy candidates with quiescent spectral and photometric properties 
at $z \gtrsim 1$ (e.g., \citealt{Bezanson2019, Bugiani2025}), and even up to $z > 3$ (e.g., \citealt{Belli2014, Glazebrook2017, Schreiber2018a_J&H_FAST, Forrest2020a, Forrest2020b}). With the advent of JWST, similar quiescent candidates have been reported at even higher redshifts (e.g., \citealt{Carnall2023, Stawinski2024, Jin2024, Nanayakkara2025, Graaff2025,  Bugiani2025}).
If these systems are genuinely quiescent at $z > 3$, their existence would imply that rapid mass assembly and efficient quenching occurred within the first 1-2 billion years of cosmic time, posing a significant challenge to current models of massive galaxy formation.

Most high-redshift candidates for massive quiescent galaxies (MQGs) have been identified primarily through near-infrared (NIR) spectroscopy and/or ultraviolet (UV)-NIR photometric surveys \citep[e.g.,][]{Schreiber2018b_UVJ, Marsan2022, Carnall2023, Forrest2020b, Forrest2024b}. 
This selection method has difficulty distinguishing truly quiescent systems from red, dust-obscured star-forming galaxies (SFGs), particularly at high redshift, especially for $z\gtrsim3$, where rest-frame optical features shift into the NIR \citep{Muzzin2013a, Hwang2021}. At these epochs, even  Spitzer/MIPS 24 \um\ and JWST/MIRI probe rest-frame wavelengths that are too blue to trace thermal dust emission, leaving the possibility of contamination from dusty galaxies uncertain.
Far-infrared (FIR) dust emission, such as that traced by ALMA, have revealed several such misclassifications, including the blended ``Jekyll and Hyde'' system at $z=3.7$ \citep{Schreiber2018a_J&H_FAST} and the dust-obscured case reported by \citet{Chang2025arXiv}, illustrating that without high-resolution far-infrared data, the nature of many red galaxies at $z>3$ remains ambiguous.



A multi-year Keck/MOSFIRE campaign, the ``Massive Ancient Galaxies At $z >$ 3 NEar-infrared'' (MAGAZ3NE) survey (\citealt{Forrest2020a, Forrest2024b}), has been successful in spectroscopically confirming a sample of ultramassive galaxies (UMGs; $\mstar > 10^{11}~\msun$) with \mbox{$3\lesssim$~\zphot~$<4$} \citep{Saracco2020, Forrest2022, Forrest2024b, McConachie2022, McConachie2025, Chang2025arXiv}.
Many MAGAZ3NE UMGs exhibit red spectral energy distributions (SEDs), and a subset show absorption features consistent with quiescence \citep{Forrest2020a,Forrest2020b}. The MAGAZ3NE spectroscopic sample has been characterized with multi-passband UV-to-NIR photometry and rest-frame optical spectroscopy.
However, at such high redshifts, dust obscuration can suppress optical emission-line indicators of star formation, such as \OII, by reprocessing ionizing radiation from young stars into thermal emission in the rest-frame far-infrared (FIR). High-resolution ALMA observations are therefore crucial to definitively rule out the possibility of dust-obscured star formation in these sources and to mitigate potential photometric contamination from nearby sources \citep{Schreiber2018a_J&H_FAST}.

Recent ALMA studies have revealed a large diversity in the dust content of MQGs at $1 < z < 4$. Several galaxies exhibit exhibit substantial dust reservoirs relative to their stellar mass despite very low specific star-formation rates, implying that quenching can proceed inefficiently, with extended dust depletion timescales \citep{Gobat2018, Suzuki2022, Lee2024}.
Conversely, other galaxies are undetected even in deep continuum observations, suggesting efficient removal or destruction of dust \citep{Whitaker2021, Gobat2022, Spilker2025arXiv}. The wide range of observed dust fractions may reflect a divergence in different quenching pathways or timescales, depending on whether the interstellar medium (ISM) is expelled by feedback or gradually consumed by residual star formation.
In this work, we use ALMA Band 7 continuum imaging to study five spectroscopically confirmed UMGs from the MAGAZ3NE survey \citep{Forrest2020b} that exhibit red UV-to-NIR SEDs and weak spectra features. The observations can provide direct measurements or upper limits on dust emission, breaking the degeneracy between truly quiescent and dust-obscured populations. It allows us to constrain the dust fraction and residual star formation, providing insight into the physical processes that drive massive galaxy quenching in the early Universe.

Our work is organized as follows. Section \ref{sec: data} summarizes the ALMA observations, data reduction, and the multi-wavelength photometric catalog. In Section \ref{sec: SED}, we describe the SED fitting procedures and the derivation of stellar properties. Section \ref{sec: QGs} examines whether the observed UMGs are truly quiescent. In Section \ref{sec: Dust}, we discuss the inferred dust properties and their implications for quenching mechanisms. A summary of our conclusions is given in Section \ref{sec: conc}.
Throughout this work, we assume a \citet{Chabrier2003} initial mass function (IMF) and a $\Lambda$CDM cosmology with $H_0=70$~km~s$^{-1}$~Mpc$^{-1}$, $\Omega_M=0.3$, and $\Omega_\Lambda=0.7$. Finally, we utilize the AB magnitude system \citep{Oke1983}.

\section{DATA}
\label{sec: data}

\subsection{Target Selection}
\label{subsec: redshift}

We selected ALMA targets from the MAGAZ3NE survey \citep{Forrest2020b, Forrest2024b}, which provided the spectroscopic follow-up sample of candidate high-redshift ultramassive galaxies at \zphot\ $\geq$ 3 with a stellar mass of \mbox{\logM~$>11.0$}. 
\citet{Forrest2020b} presented spectroscopically confirmed absorption-line UMGs at \zspec\ $> 3$ with UV-to-NIR SED-inferred star formation rates of at most a few $M_{\odot}~\mathrm{yr^{-1}}$ (see details in Section 4 of \citealt{Forrest2020b}). 
The ALMA observations presented here target the four most quiescent systems from this population, selected for their particularly low inferred star formation activity. These galaxies potentially experienced intense star formation episodes and rapid quenching.
Four galaxies were observed with Keck/MOSFIRE in the $H$- band and $K$-band, with the exception of XMM-VID1-2075, which has only $K$-band coverage, targeting the expected positions of the \Hbeta\ and \OIIshort\ emission lines.


\begin{figure*}
\centering
\includegraphics[width = 0.98\textwidth]{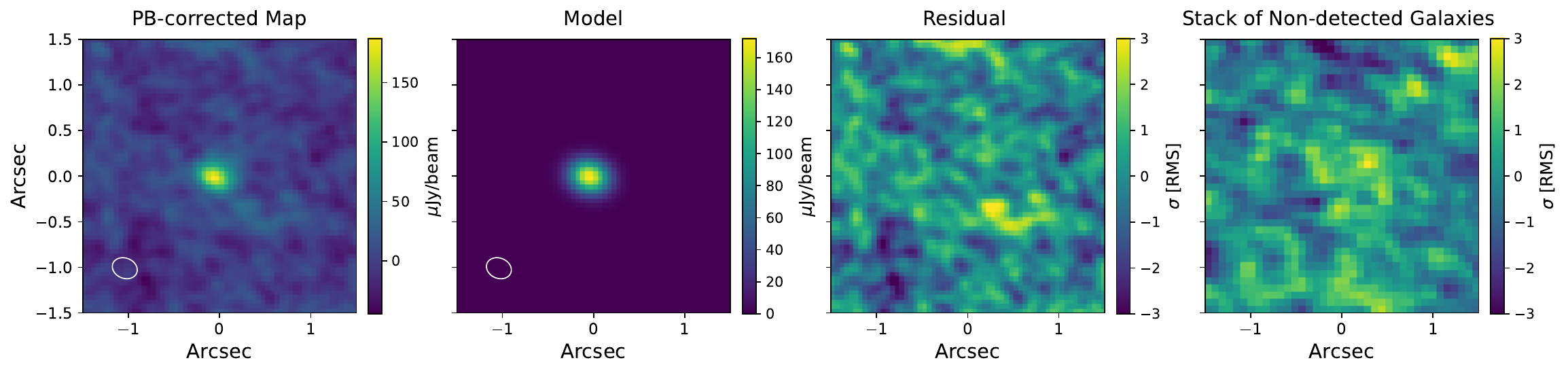}
\caption{From left to right: ALMA Band 7 (870 $\mu$m) primary-beam-corrected continuum image of XMM-VID3-2457; the two-dimensional Gaussian model; the residual map after subtraction of the model, shown in units of the RMS; and the stacked continuum image of four non-detected galaxies.
The 0.28\arcsec $\times$ 0.22\arcsec synthesized beam of XMM-VID3-2457 is represented by the white ellipse at the bottom left corners. 
\label{fig: ALMA_cont}}
\end{figure*}


\subsection{ALMA Observation and Reduction}
\label{subsec: ALMA_Reduction}

\begin{deluxetable*}{ccccccc}
\tabletypesize{\small}   
\centering
\caption{ALMA band 7 (872 $\mu$m) observation parameters. Flux density is estimated from pb-corrected images using \texttt{imfit} task. 
\label{tab: ALMA_DATAreduction}}
\tablehead{
\textbf{Object} & \textbf{z$_{\rm spec}$} & \textbf{Int. Time} & \textbf{Flux}   & \textbf{RMS}    & 
\textbf{Beam Size} & \textbf{QA2 status}\\
             &  & (s)   & ($\mu$Jy)           & ($\mu$Jy beam$^{-1}$)              &     & \\}
\startdata
XMM-VID3-2457          & 3.498$^{+0.003}_{-0.002}$ & 8437 & $294 \pm 33$    & 12.6             & 
0.28$'' \times$ 0.22$''$, PA = 68$^\circ$ & PASS \\
XMM-VID3-1120           & 3.492$^{+0.002}_{-0.003}$ & 8437 &  $<35.1$ ($3\sigma$)     & 11.7       & 0.28$'' \times$ 0.22$''$, PA = 68$^\circ$ & PASS\\
XMM-VID1-2075         & 3.495$^{+0.001}_{-0.002}$ & 8437 &  $<35.4$ ($3\sigma$)    & 11.8     & 0.28$'' \times$ 0.22$''$, PA = 68$^\circ$  & PASS\\
COS-DR3-201999      &3.131$^{+0.001}_{-0.001}$   & 2873 &  $<55.8$ ($3\sigma$)     & 18.6  & 0.42$'' \times$ 0.29$''$, PA = 70$^\circ$ & SEMIPASS \\
COS-DR3-202019\tablenotemark{a}    &3.133$^{+0.002}_{-0.001}$    & 2873 &  $<55.8$ ($3\sigma$)     & 18.6           
& 0.42$'' \times$ 0.29$''$, PA = 70$^\circ$ 
& SEMIPASS\\
\enddata
\tablenotetext{a}{COS-DR3-202019 lies within the primary beam of the COS-DR3-201999 pointing.}
\end{deluxetable*}

Band 7 (872 $\mu$m) continuum observations of four targeted sources were carried out in the Cycle 8 program (Project ID: 2021.1.00501.S, PI: B. Forrest). 
We obtained the calibrated measurement sets (MSs) from the National Radio Astronomy Observatory (NRAO), which were calibrated using Pipeline version 2021.2.0.128.\footnote{Download MS from NRAO: https://data.nrao.edu/portal/} 
Three XMM-VIDEO targets were observed using the ALMA 12-meter array, which has 61 antennas and successfully passed the QA2 quality level. The scheduling block for target COS-DR3-201999 timed out due to the end of the observation cycle. This target was observed using the 12-meter array with 42 antennas. 
In addition to the four UMGs selected for ALMA observations, the ALMA field of view (FoV) includes an additional MAGAZ3NE UMG in the COSMOS field, COS-DR3-202019 (\zspec\ = 3.133), located 7.4\arcsec\ from COS-DR3-201999 (\zspec\ = 3.131).  
This source was previously identified in \citet{Forrest2020b} as an \OIIshort\ emitting, star-forming galaxy based on its rest-frame optical spectra and UV-NIR photometry. Given its inclusion in the FoV, we incorporate it into our analysis to investigate its dust properties. In total, this work analyzes five UMGs, including the four originally selected ALMA targets and COS-DR3-202019.

We further processed and reduced the data for all sources with the version 6.6.3.22 package of Common Astronomy Software Application (\texttt{CASA}). The calibrated visibilities were imaged using the \texttt{TCLEAN} algorithm in \texttt{CASA}, with natural weighting (ROBUST = 2) and pixel scales of 0.05\arcsec\ for XMM targets and 0.08\arcsec\ for COSMOS targets. 
The images were cleaned to the Root Mean Square (RMS) noise of 11.7$-$18.6 $\mu$Jy beam$^{\rm -1}$. Three XMM targets achieve an RMS noise level consistent with the expected 14 $\mu$Jy beam$^{\rm -1}$, while the COSMOS observations are slightly less sensitive due to the reduced exposure time.
We then applied task \texttt{imfit} in \texttt{CASA} to measure the integrated fluxes for all UMGs and deconvolved beam sizes (FWHM) through a two-dimensional (2D) Gaussian fitting procedure. 
Only UMG XMM-VID3-2457 is detected in Band 7 continuum with a flux of 294 $\pm$ 33 $\mu$Jy, while the remaining four targets are undetected down to the RMS sensitivity.

Figure~\ref{fig: ALMA_cont} shows the 872 \um\ continuum emission for the only detected source, XMM-VID3-2457, together with the best-fit 2D Gaussian model and the corresponding residual map.
No region exhibits a significant excess beyond 3$\sigma$.
The remaining four UMGs are individually undetected at the $3\sigma$ level in the ALMA continuum maps. In order to verify the absence of significant emission, we construct a stacked continuum image after convolving all maps to the lowest angular resolution in the sample, that of COS-DR3-201999.
The stacking is performed using inverse-variance weighting based on the local RMS noise measured in the primary-beam-uncorrected images.
The resulting stacked image is shown in the right panel of Figure\ \ref{fig: ALMA_cont}, reaches an rms sensitivity of $\sigma_{\rm stack, RMS} = 8.0$ $\mu \rm Jy$ $\rm beam^{-1}$, and does not reveal any significant continuum emission. 
Table \ref{tab: ALMA_DATAreduction} provides the observational parameters and the processed data for five UMGs.  
During the SED fitting in Section\ \ref{sec: SED}, we adopt 3$\sigma$ upper limits on the ALMA band 7 photometry for the UMGs without detections.




\subsection{UV-to-NIR Photometry}
\label{subsec: photometry}

 
Three ultramassive galaxies were selected in the XMM-Newton Large Scale Structure field from the VISTA Deep Extragalactic Observations survey (VIDEO; \citealt{Jarvis2013}). This field was divided into three VIDEO survey tiles (XMM-VID1, VID2, and VID3), distinguishing sources by their location within the survey. The photometric catalog was constructed using VIDEO DR4 data over 4.65 deg$^{2}$ with a 5 $\sigma$ depth of K$_{s}$ = 23.8 mag (\citealt{Forrest2020b}; Annunziatella et al. 2026, in prep.). These contain 22 photometric passbands ranging from $u$ band to IRAC 8.0 $\mu$m, including deep near-IR observations at $Z$-, $Y$-, $H$-, and $K_{s}$-band 
from the VIDEO survey \citep{Jarvis2013}, deep IRAC data from SERVS \citep{Mauduit2012} and the Deep Drill survey \citep{Lacy2021}. 

Two additional UMGs were selected from the UltraVISTA near-infrared imaging survey \citep{McCracken2012}, which covers the COSMOS field \citep{Scoville2007}. 
The $K_{s}$-selected photometric catalog (\citealt{Marsan2022}) was constructed with \textit{ultra-deep} UltraVISTA imaging Data Release 3 (hereafter DR3) covering 0.84 deg$^{2}$ and reaching a depth of 5$\sigma$ of $K_{s}$ = 25.2 mag within a 2{\farcs}1 diameter aperture, following the same methodology as described in \citet{Muzzin2013a}. The DR3 catalog provides ancillary photometry spanning 0.15-24 \um\ across 49 bands, with ``COS-DR3" denoting sources from the UltraVISTA DR3 catalog.


\begin{figure}
\centering
\includegraphics[width = 0.49\textwidth]{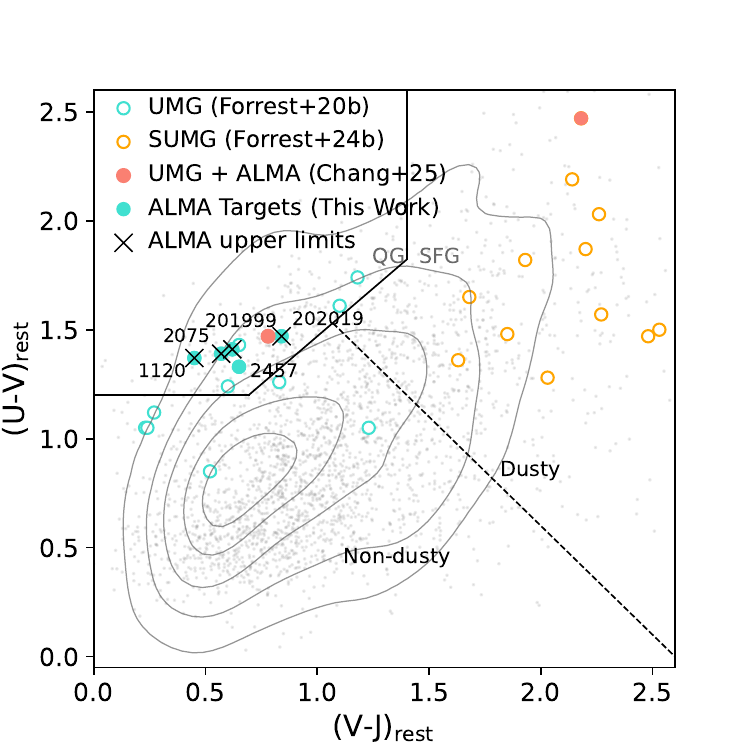}
\caption{
Rest-frame (RF) \UVJ\ color-color diagram for 16 spectroscopically-confirmed UMG candidates (open cyan circles) presented in \cite{Forrest2020b} and 12 spectroscopically-confirmed S-UMGs candidates (open orange circles) presented in \cite{Forrest2024b}.  
The five UMGs from \cite{Forrest2020b} observed with ALMA in this work are shown by filled cyan circles (Galaxy 1: XMM-VID3-2457; 2: XMM-VID3-1120; 3: XMM-VID1-2075; 4: COS-DR3-201999; 5: COS-DR3-202019), where cross symbols indicate non-detections in ALMA dust continuum with upper limits only.
Two additional ALMA-detected UMGs analyzed previously in \citet{Chang2025arXiv} are shown as filled salmon circles.
The RF colors were estimated at the spectroscopic redshift using FAST++ and UV-to-8\um\ photometry only.
The dashed line denotes the division between  ``dusty'' and ``non-dusty'' star-forming regions \citep{Schreiber2018b_UVJ}. 
The gray data points corresponds to massive galaxies with \mbox{\logM\ $> 10.5$} at \mbox{$3 < z_{\rm phot} < 4$} in the COSMOS2020 catalog \citep{Weaver2022}, with 
their density distribution overplotted as gray contours. 
\label{fig: UVJ}}
\end{figure}

Figure\ \ref{fig: UVJ} shows the rest-frame \textit{U-V} and \textit{V-J} (RF \UVJ; hereafter) colors of the five UMGs, together with other spectroscopically-confirmed MAGAZ3NE UMGs \citep{Forrest2020b} and Super-UMG candidates ($M_{\rm phot} > 10^{11.7}~{\rm M}_{\odot}$; \citealt{Forrest2024b}). The RF colors were derived from UV-to-8$\mu$m photometry fit with FAST$++$ at the spectroscopic redshift of each galaxy, where the $K$-band flux was corrected for emission line fluxes (for full details, see \citealt{Forrest2020b}). 
As shown in the figure, all five UMGs lie securely within the quiescent wedge of the $UVJ$ diagram \citep{2011Whitaker}.
This placement suggests that they are consistent with quiescent population, rather than dusty star-forming galaxies.
%
Our comparison with massive galaxies with \logM\ $> 10.5$ at \(3 < z_{\rm phot} < 4\) from the COSMOS2020 catalog \citep{Weaver2022} shows no clear bimodality in Figure\ \ref{fig: UVJ}.

\subsection{Ancillary Photometry in the COSMOS Field}
\label{subsec: Jin18}

For the two UMGs in this work that lie in the COSMOS field, which benefit from extensive multi-wavelength coverage, we complement the UltraVISTA DR3 photometric catalog with an IR-to-radio deblended photometric catalog \citep[][hereafter as {Jin18}]{Jin2018}. Detailed cross-matching between these catalogs is described in \citet{Chang2025arXiv}.
The Jin18 catalog employs prior-based deblending techniques to mitigate the effects of source confusion and blending caused by the large beam sizes of FIR observations. 
The two COSMOS UMGs in this work are matched to sources in the Jin18 catalog using their DR3 coordinates, with a positional offset $\leq$ 0.2\arcsec.
We adopt the \textit{Spitzer}/MIPS 24 \um\ (PI: D. Danders; \citealt{LeFloch2009}) and VLA 3 GHz and 1.4 GHz \citep{Smolvcic2017, Schinnerer2010} photometry from Jin18. \textit{Herschel} and SCUBA2 data are excluded because the environments of COS-DR3-202019 and COS-DR3-20199 are highly crowded, with over ten bright $K_s$-band sources within 10\arcsec, well below the \textit{Herschel} (FWHM $\sim7$\arcsec -36\arcsec) and SCUBA2 850 \um\ (FWHM $\sim$11\arcsec) beam sizes.
For COS-DR3-201999, no radio counterpart is reported in the VLA catalog of \citet{Smolvcic2017}. To avoid adopting a potentially uncertain deblended flux, we instead derive a radio upper limit directly from the VLA 3 GHz and 1.4GHz images. The RMS noises are measured in source-free regions near the target using Cube Analysis and Rendering Tool for Astronomy (\texttt{CARTA}), and the resulting value is used to define the corresponding (3$\sigma$) upper limit.

%

\begin{figure*}[!t]
\centering
\includegraphics[width = 0.21\textwidth]{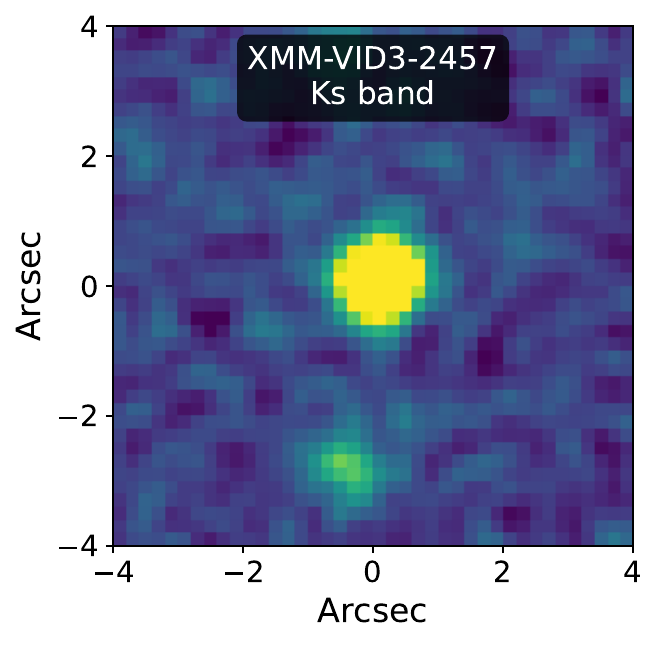}
\includegraphics[width = 0.21\textwidth]{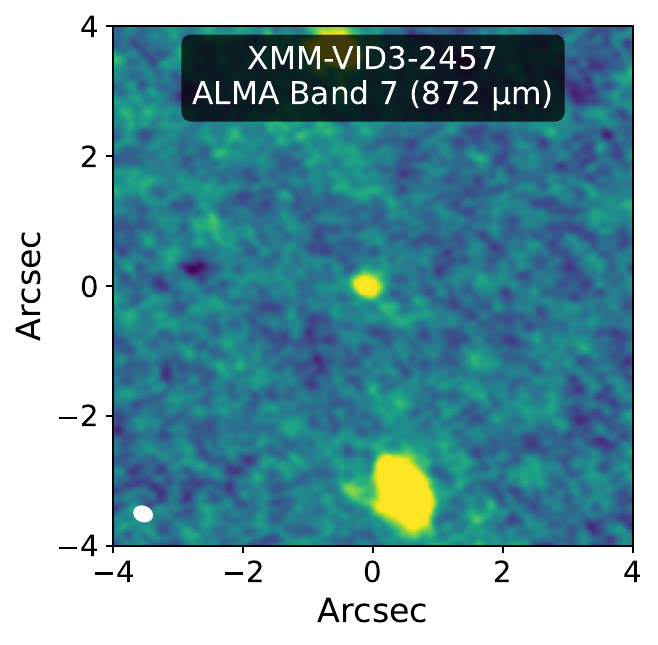}
\includegraphics[width = 0.55\textwidth]{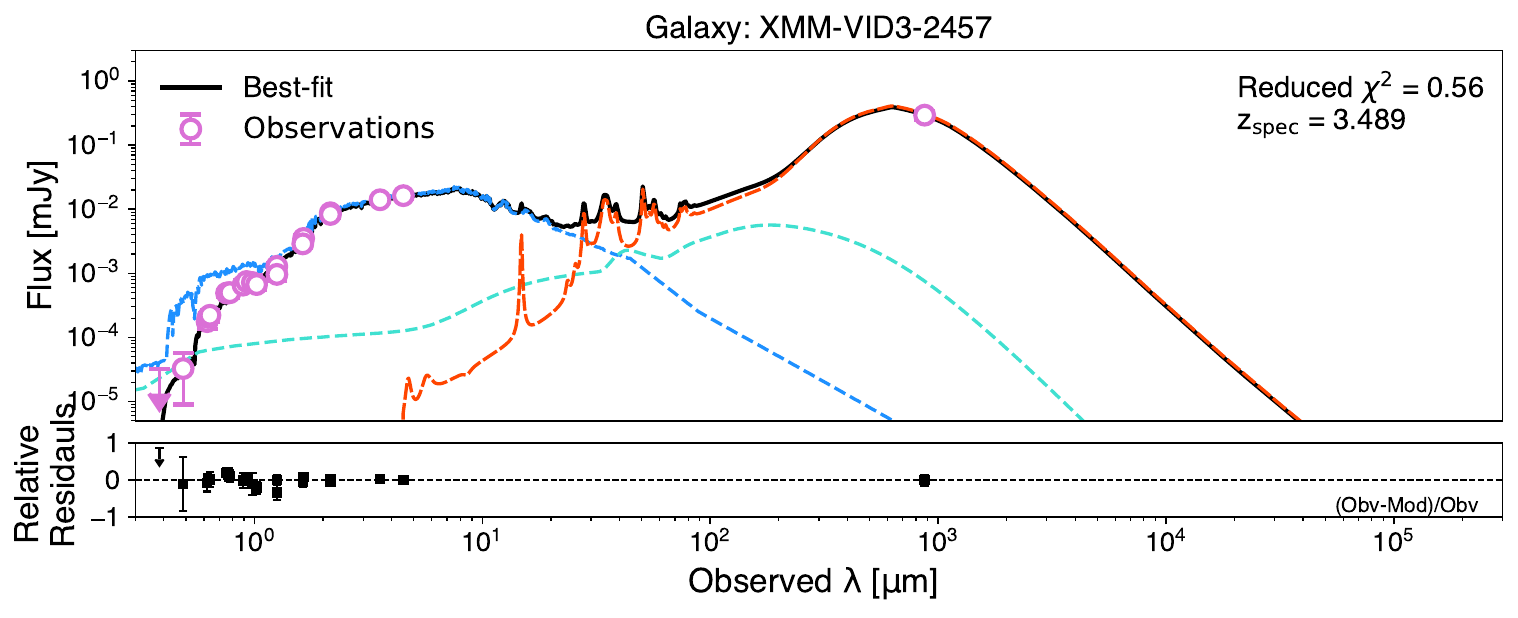}
\includegraphics[width = 0.21\textwidth]{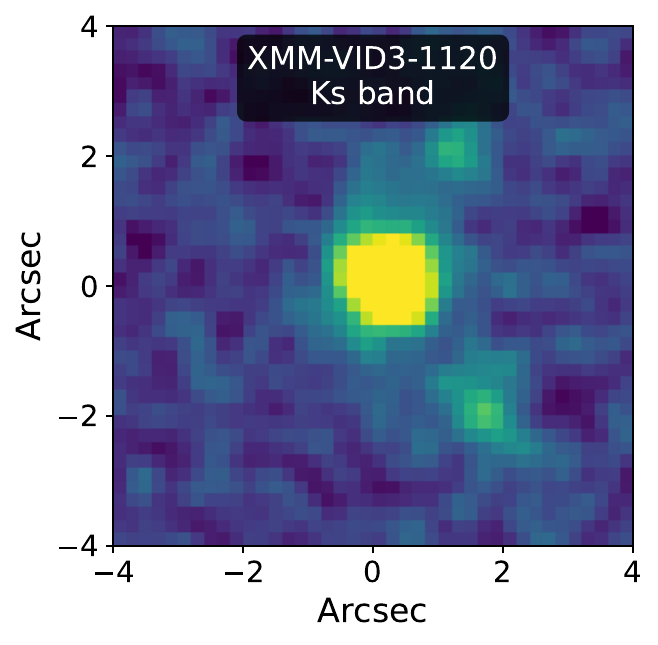}
\includegraphics[width = 0.21\textwidth]{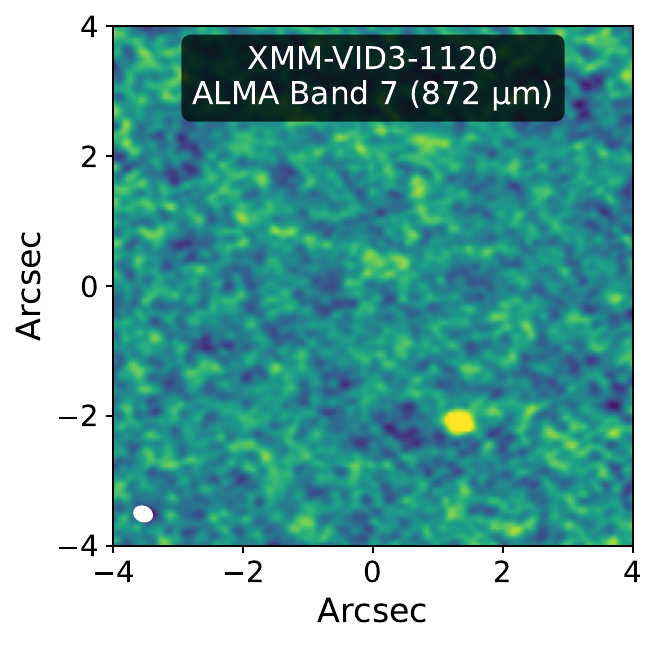}
\includegraphics[width = 0.55\textwidth]{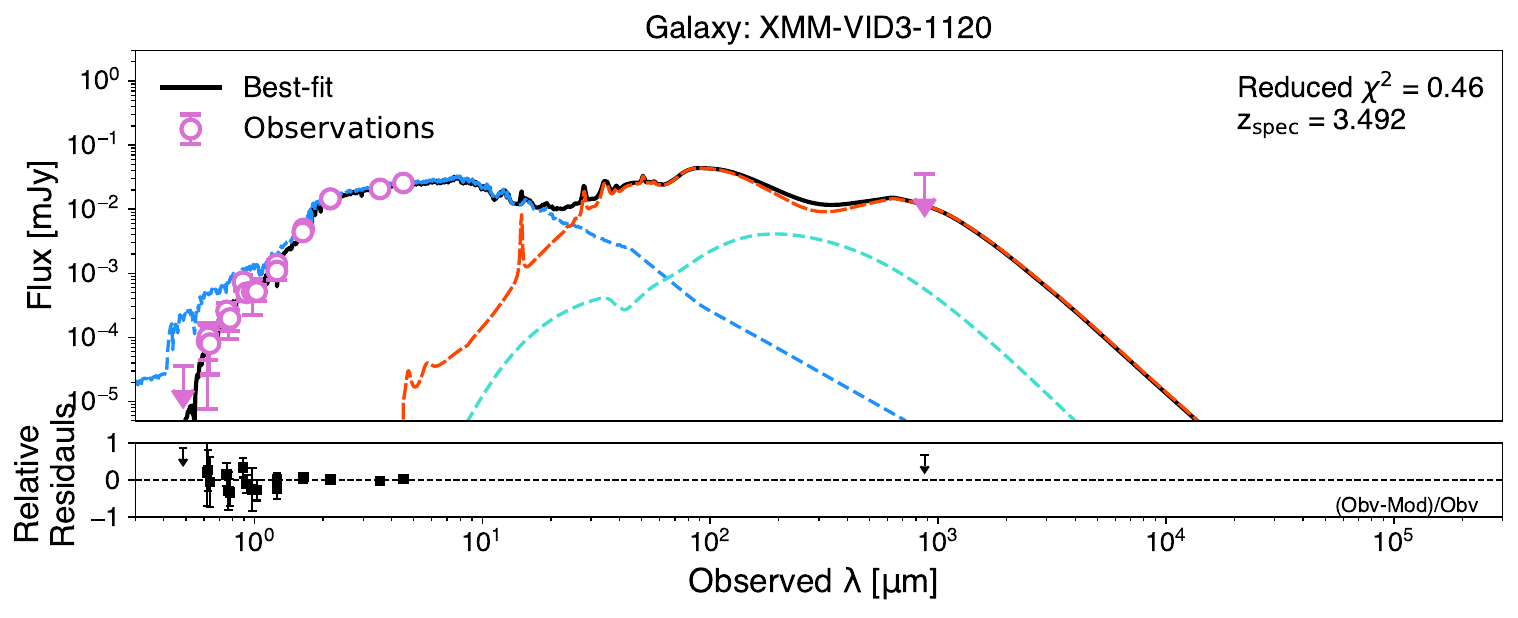}
\includegraphics[width = 0.21\textwidth]{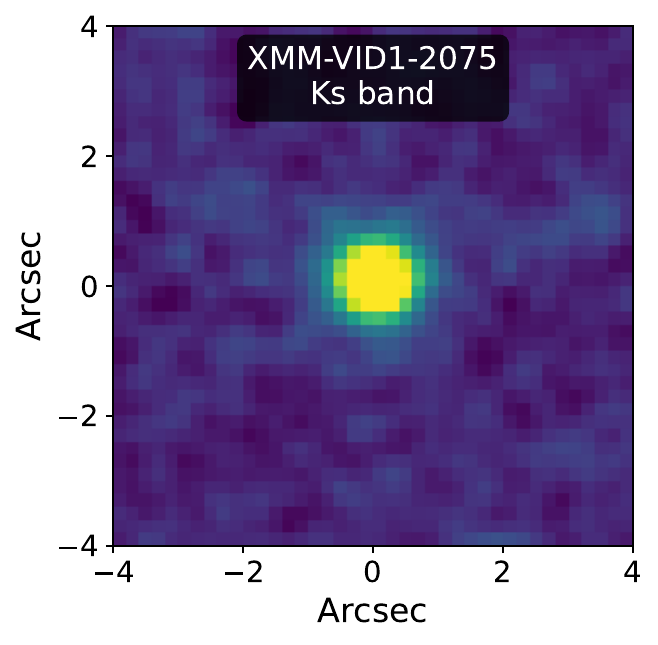}
\includegraphics[width = 0.21\textwidth]{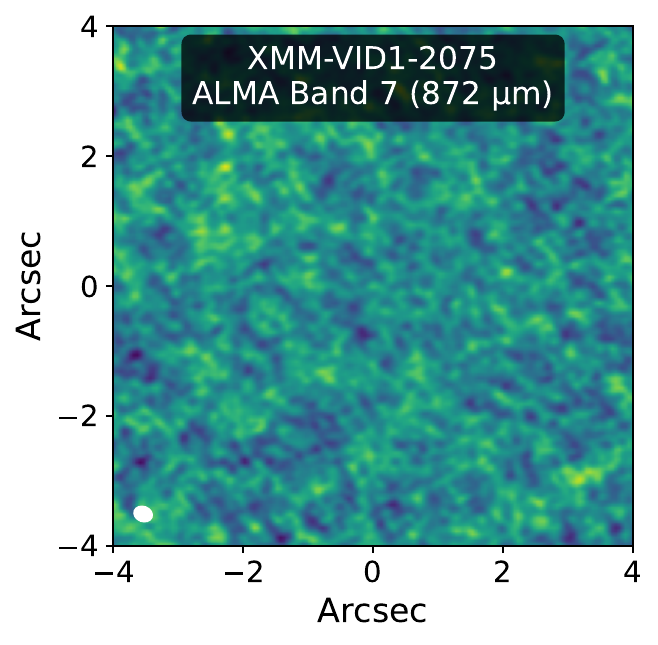}
\includegraphics[width = 0.55\textwidth]{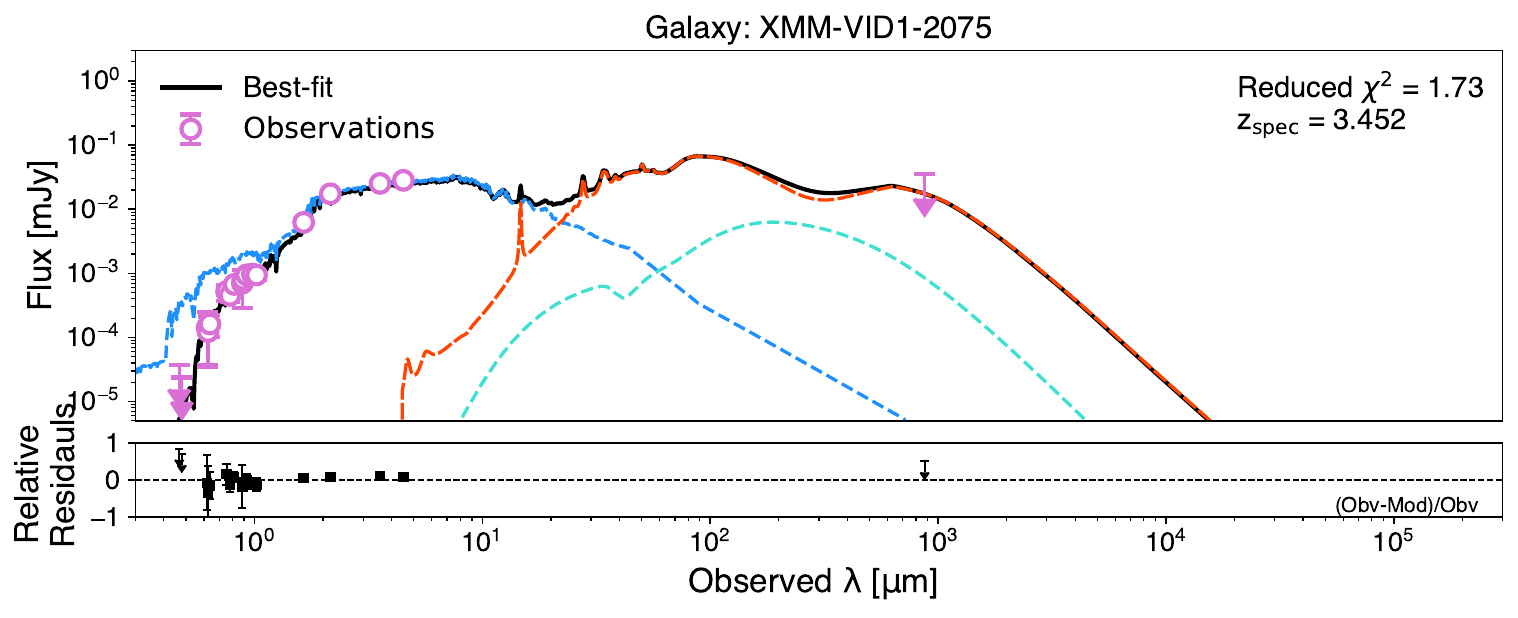}
\includegraphics[width = 0.21\textwidth]{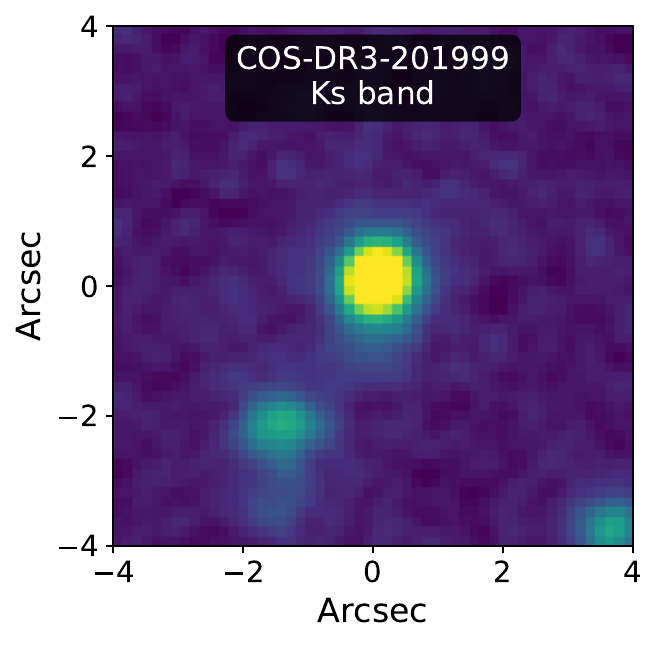}
\includegraphics[width = 0.21\textwidth]{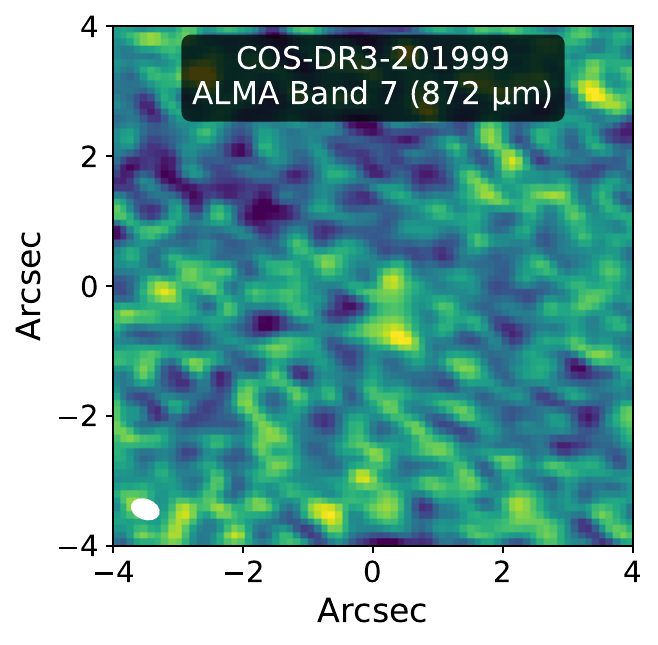}
\includegraphics[width = 0.55\textwidth]{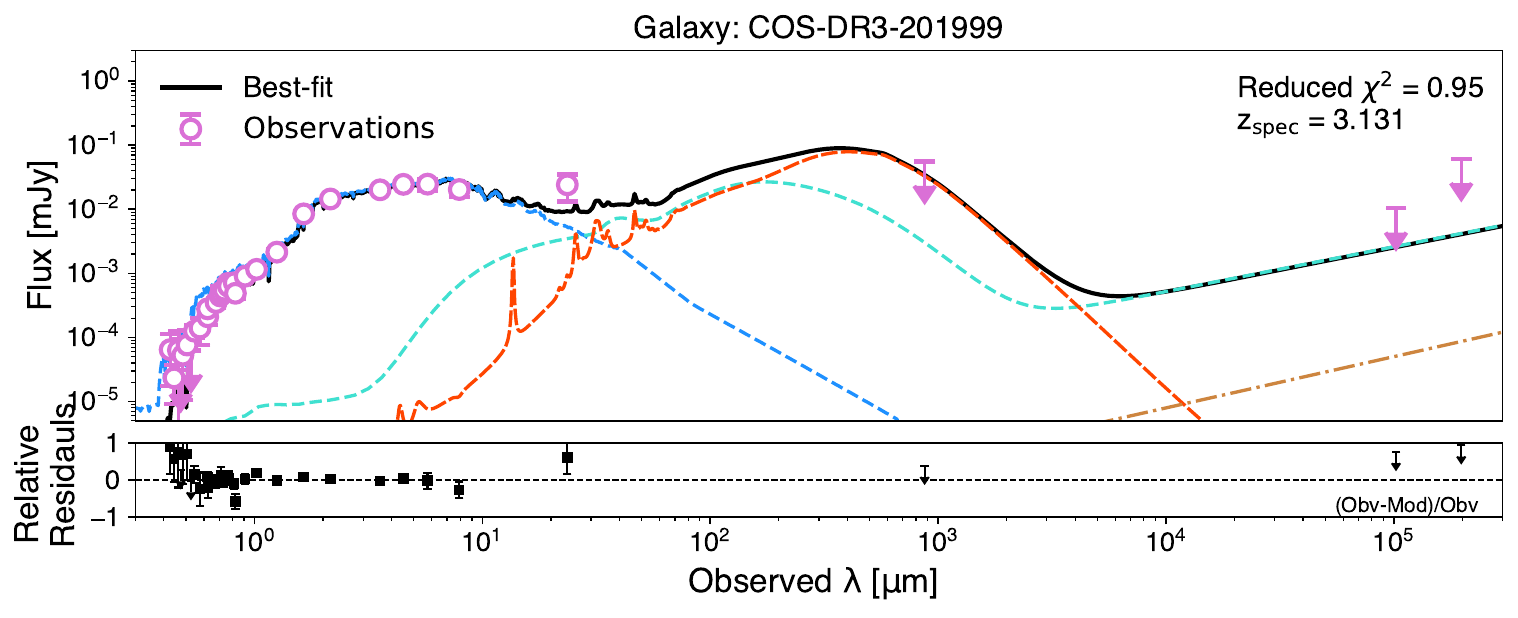}
\includegraphics[width = 0.21\textwidth]{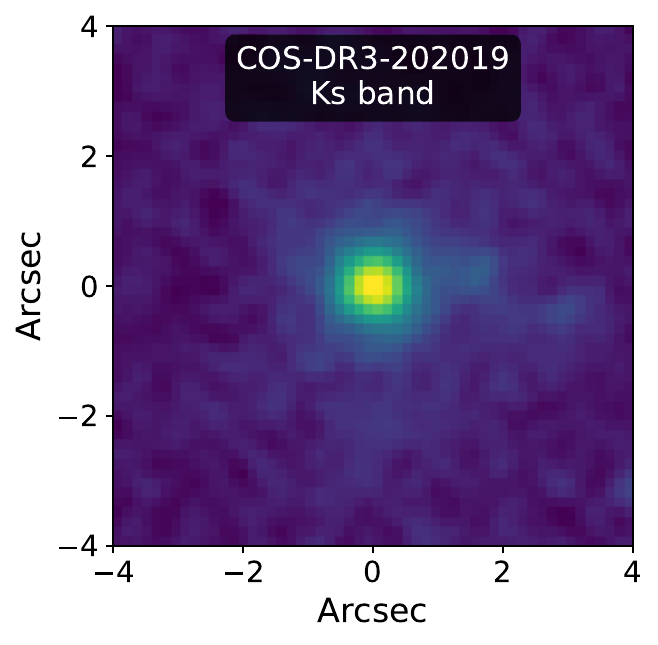}
\includegraphics[width = 0.21\textwidth]{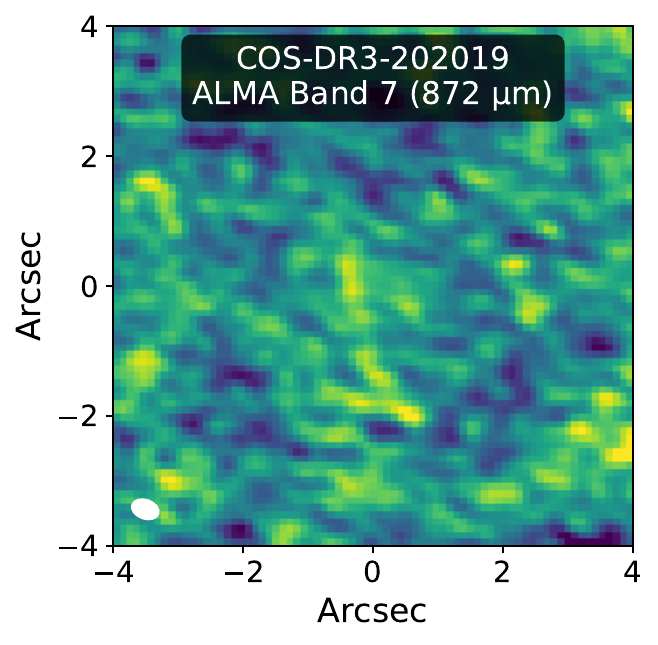}
\includegraphics[width = 0.55\textwidth]{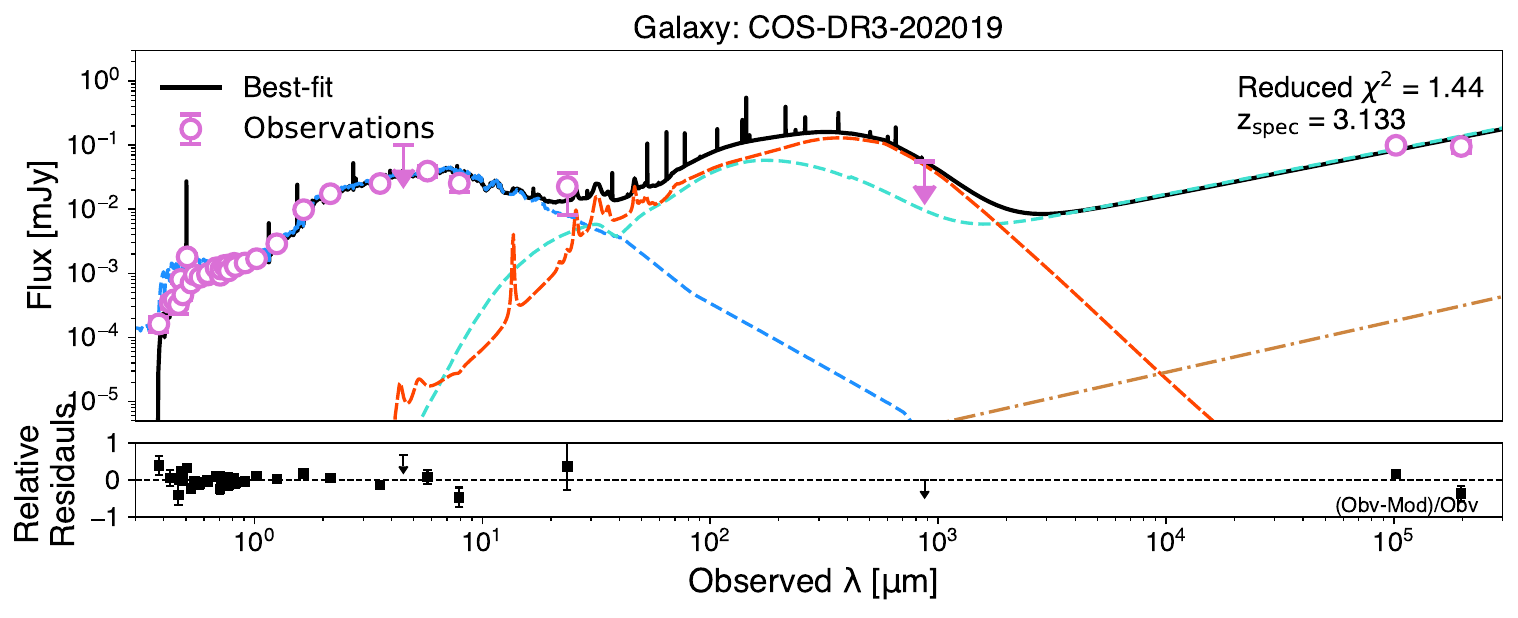} 
\caption{$K_s$-band images, ALMA band 7 (872\um) primary-beam-corrected continuum maps, and UV-to-millimeter/radio SED modeling of the five UMGs. 
Imaging cutouts are centered on the spectroscopic positions of the galaxies. 
White ellipses in the lower-left corners of the ALMA maps show the synthesized beam. 
The SED panels on the right show the best-fit model (top) and relative residuals (bottom). 
Dashed lines indicate the individual \texttt{CIGALE} SED components: unattenuated stellar emission (blue), dust emission (red), and AGN emission (cyan). For UMGs COS-DR3-201999 and COS-DR3-202019, an additional synchrotron radio emission component is included (brown). 
Observed photometry is shown as purple circles, with downward arrows indicating upper limits. 
\label{fig: SED_all}}
\end{figure*}

\renewcommand{\arraystretch}{1.2} 
\setlength{\tabcolsep}{4.5pt}     
\begin{deluxetable*}{lcccccc}
\tabletypesize{\footnotesize}   
\tablecaption{Physical properties for five UMGs. 
\label{tab:prop_all}} 
\tablehead{
\colhead{ObjID} & \colhead{Method} & \colhead{XMM-VID3-2457} & \colhead{XMM-VID3-1120} & \colhead{XMM-VID1-2075}  & \colhead{COS-DR3-201999} & \colhead{COS-DR3-202019}
}
\startdata  
redshift   & \texttt{EAZY} &  $3.51^{+0.07}_{-0.07}$   & $3.40^{+0.12}_{-0.10}$   & $3.48^{+0.08}_{-0.07}$   & $3.14^{+0.09}_{-0.09}$   & $3.10^{+0.06}_{-0.04}$ \\
  & \texttt{slinefit} & $3.498^{+0.003}_{-0.002}$ & $3.492^{+0.002}_{-0.003}$ & $3.452^{+0.001}_{-0.002}$ & $3.131^{+0.001}_{-0.001}$ & $3.133^{+0.002}_{-0.001}$ \\
\hline
\multicolumn{7}{c}{\textbf{Stellar Properties}} \\
\hline
$\rm \log M_{\star} [M_{\odot}]$ &  \texttt{F20b} & $11.26^{+0.02}_{-0.03}$ & $11.47^{+0.02}_{-0.03}$ & $11.52^{+0.00}_{-0.05}$ & $11.40^{+0.03}_{-0.01}$ & $11.67^{+0.04}_{-0.05}$ \\ 
   &  \texttt{CIGALE} & $11.25^{+0.06}_{-0.07}$ & $11.51^{+0.02}_{-0.02}$ & $11.50^{+0.03}_{-0.03}$ & $11.33^{+0.02}_{-0.03}$ & $11.49^{+0.05}_{-0.05}$ \\ 
\hline
SFR [$\rm M_\odot$ yr$^{-1}$]       &  \texttt{F20b} & $< 0.4$  & $< 0.1 $  & $<1.0$  & $1.3^{+0.3}_{-1.2} $  & $82.4^{+10.4}_{-58.2} $ \\
   &  \texttt{CIGALE} & $8.7 \pm 4.1$  & $4.1 \pm 2.2$  & $4.9 \pm 2.8$  & $1.1 \pm 0.7$  & $18.1 \pm 2.1$ \\
  & \texttt{$L_{\rm IR}$} & $12.6 \pm 4.1$ & $8.4 \pm 3.0$ & $9.7 \pm 4.3$ & $3.8 \pm 0.6$ & $7.4 \pm 1.1$ \\
  & H$\beta$ & $2.2 \pm 3.8$ & $4.2 \pm 4.9$ & $9.1 \pm 2.2$ & $1.5 \pm 2.5$ & $36.5 \pm 13.3$ \\
  &  [O II] & $2.5 \pm 1.9$ & $2.8 \pm 2.3$ & $-$ & $6.3 \pm 1.2$ & $57.6 \pm 2.6$ \\
\hline
$\log$ sSFR [yr$^{-1}$]         &  \texttt{F20b} & $<-11.63$ & $<-12.44$ & $<-11.47$ & $-11.29^{+0.09}_{-1.11}$ & $-9.75^{+0.07}_{-0.53}$ \\
     &  \texttt{CIGALE} & $-10.31^{+0.23}_{-0.33}$ & $-10.90^{+0.21}_{-0.35}$ & $-10.81^{+0.23}_{-0.39}$ & $-11.27^{+0.23}_{-0.42}$ & $-10.24^{+0.10}_{-0.10}$ \\
\hline
A$_{\rm V}$  &  \texttt{F20b} & $0.5 \pm 0.02$  & $0.0 \pm 0.0$  & $0.4 \pm 0.2$  & $0.4 \pm 0.1$  & $0.6 \pm 0.3$ \\
  &  \texttt{CIGALE} & $0.10 \pm 0.05$  & $0.05 \pm 0.02$  & $0.04 \pm 0.02$  & $0.03 \pm 0.01$  & $0.03 \pm 0.01$ \\
\hline
frac$_{\rm AGN}$                &  \texttt{CIGALE} & $0.16 \pm 0.11$  & $0.14 \pm 0.09$  & $0.11 \pm 0.05$  & $0.50 \pm 0.10$  & $0.47 \pm 0.10$ \\
\hline
\multicolumn{7}{c}{\textbf{Dust Properties}} \\
\hline
$\rm \log L_{\rm dust} [L_{\odot}]$ &  \texttt{CIGALE} & $11.10^{+0.17}_{-0.12}$ & $10.93^{+0.13}_{-0.19}$ & $10.98^{+0.16}_{-0.25}$ & $10.58^{+0.07}_{-0.08}$ & $10.87^{+0.06}_{-0.07}$ \\
$\rm \log M_{\rm dust} [M_{\odot}]$ &  \texttt{CIGALE} & $8.43^{+0.15}_{-0.22}$ & $7.14^{+0.15}_{-0.23}$ & $7.23^{+0.13}_{-0.28}$ & $7.40^{+0.22}_{-0.48}$ & $7.42^{+0.15}_{-0.23}$ \\
 & \texttt{MBB} & $8.22^{+0.05}_{-0.05}$ & $<7.30$ & $<7.30$ & $<7.49$ & $<7.49$ \\
\enddata
\end{deluxetable*}

\section{SED Fitting} 
\label{sec: SED}

To investigate the dust content and star formation activity of these high-$z$ UMGs, we perform multi-wavelength SED fitting with \texttt{CIGALE} (Code Investigating GALaxy Emission)\footnote{CIGALE v2022.1: \url{https://cigale.lam.fr/}}, a flexible energy balance SED modeling code that relies on Bayesian inference to estimate the physical properties of galaxies \citep{Boquien2019, Yang2020, Yang2022}. In this framework, the stellar light absorbed by dust at UV-NIR wavelengths are redistributed and emitted consistently in the MIR and FIR. 
Here, we briefly describe the modules used to fit the SEDs for our sample. Full details of input parameters may be found in Table\ \ref{tab:SED_input}. 
We adopt a delayed exponentially declining star formation history (SFH) and simple stellar populations (SSPs) from \citet{BruzualCharlot03}, assuming a \citet{Chabrier2003} IMF. Nebular emission is included, parameterized by the ionization parameter and gas-phase metallicity. We employ the dust modified attenuation law of \citet{Calzetti2000} and the dust emission templates of \citet{Draine2014}. AGN emission is modeled using the clumpy \texttt{SKIRTOR} framework \citep{Stalevski2012, Stalevski2016}, including viewing representing face-on (type 1) and edge-on (type 2) AGN.
As noted in Section~\ref{subsec: Jin18}, COS-DR3-201999 and COS-DR3-202019 are detected at VLA 3 GHz and 1.4 GHz. For these two UMGs across UV-to-radio range, we additionally include the radio module to account for a power-law  synchrotron spectrum and AGN radio emission.

The best-fit SEDs and imaging of K$_s$ band and ALMA band 7 for five UMGs are presented in Figure\ \ref{fig: SED_all}, alongside a comparison with the $H$- and/or $K$-band spectra observed by Keck/MOSFIRE, which demonstrate a high level of consistency between the best-fit model and the spectroscopic data. 
The Bayesian-derived properties of the five UMGs from the \texttt{CIGALE} SED fitting are summarized in Table~\ref{tab:prop_all}, together with the stellar properties obtained using FAST$++$ \citep{fastpp} as reported in \citet{Forrest2020b}, and SFRs derived from multiple tracers. 
\citet{Forrest2020b} reported stellar masses and SFRs for MAGAZ3NE UMGs using FAST$++$ fits to the UV-NIR photometry and spectroscopy, performed without AGN templates. 
The stellar masses obtained from FAST$++$ and \texttt{CIGALE} are consistent within 0.1 dex for four of the ALMA-targeted UMGs. The only exception is COS-DR3-202019, a close companion of COS-DR3-201999, for which \texttt{CIGALE} derives a stellar mass lower by $\sim$0.25~dex. 
This offset potentially reflects the inclusion of AGN and radio components in the \texttt{CIGALE} modeling, which are not accounted for in the FAST$++$ fits, which we discuss in Section~\ref{subsec: AGN}. In the following sections, we adopt the \texttt{CIGALE}-derived results to discuss the star-formation activity and dust emission properties of these five UMGs (Sections~\ref{sec: QGs} and~\ref{sec: Dust}).



\section{Suppressed Star Formation in UMGs}
\label{sec: QGs}

\subsection{Star-Forming Main Sequence}
\label{subsec: SFMS}


\begin{figure}
\centering
\includegraphics[width = 0.48\textwidth]{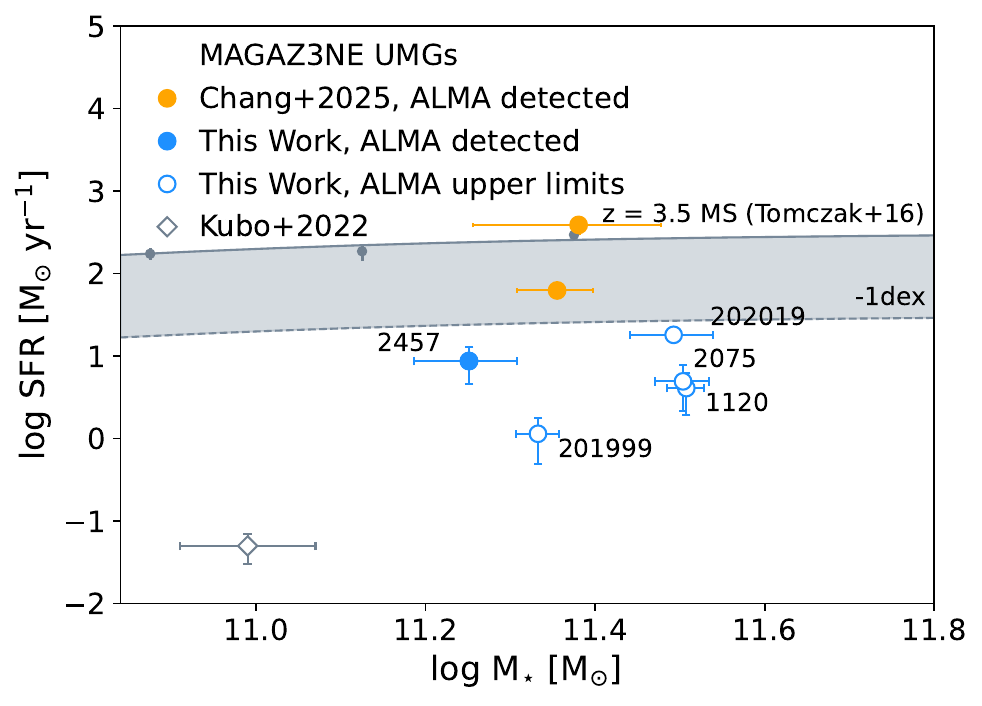}
\caption{
The SFR versus stellar mass.
Blue filled circles show the \texttt{CIGALE}-derived measurements for the five UMGs in this work, while open circles mark UMGs with ALMA upper limits only.
Gray symbol represents the spectroscopically confirmed quiescent galaxy at $3<z<4$ observed with ALMA and fitted with \texttt{CIGALE} in \citet{Kubo2022}.
The gray line shows the star-forming main sequence from \citet{Tomczak2016}. 
Shaded regions denote 1~dex below the SFMS, corresponding to the threshold for classifying galaxies as quiescent comparing to the main sequence.
Orange filled circles represent two MAGAZ3NE UMGs analyzed using UV-to-FIR \texttt{CIGALE} SED with ALMA observations in \citet{Chang2025arXiv}. 
\label{fig: SFMS}}
\end{figure}

With the inclusion of ALMA data into the UV-to-FIR SED fitting, we present the \texttt{CIGALE}-derived stellar mass and SFR$_{\rm 10Myrs}$ of five UMGs with the star-forming main sequence (SFMS) at $z = 3.5$ in Figure~\ref{fig: SFMS}. 
Filled circles represent UV-to-FIR SED fits from this work, while orange filled circles indicate the two MAGAZ3NE UMGs (COS-DR3-195616 and COS-DR1-209435) previously analyzed with UV-to-FIR SED fitting by \citet{Chang2025arXiv} including ALMA observations.
All five UMGs lie significantly ($>$ 1 dex) below the SFMS (\mbox{$\Delta \mathrm{\log SFR} < -1$}) from \citet{Tomczak2016}, classifying them as quiescent systems at $z>3$. 
Figure~\ref{fig: SFMS} compares the SED-derived quantities of other spectroscopically confirmed quiescent galaxies at $3<z<4$ observed with ALMA presented by \citet{Kubo2022} fitted with \texttt{CIGALE}.
The five MAGAZ3NE UMGs occupy the ultramassive regime (\logM\ $>11.2$), where spectroscopically confirmed systems remain rare. 
   

\begin{figure}
\centering
\includegraphics[width = 0.48\textwidth]{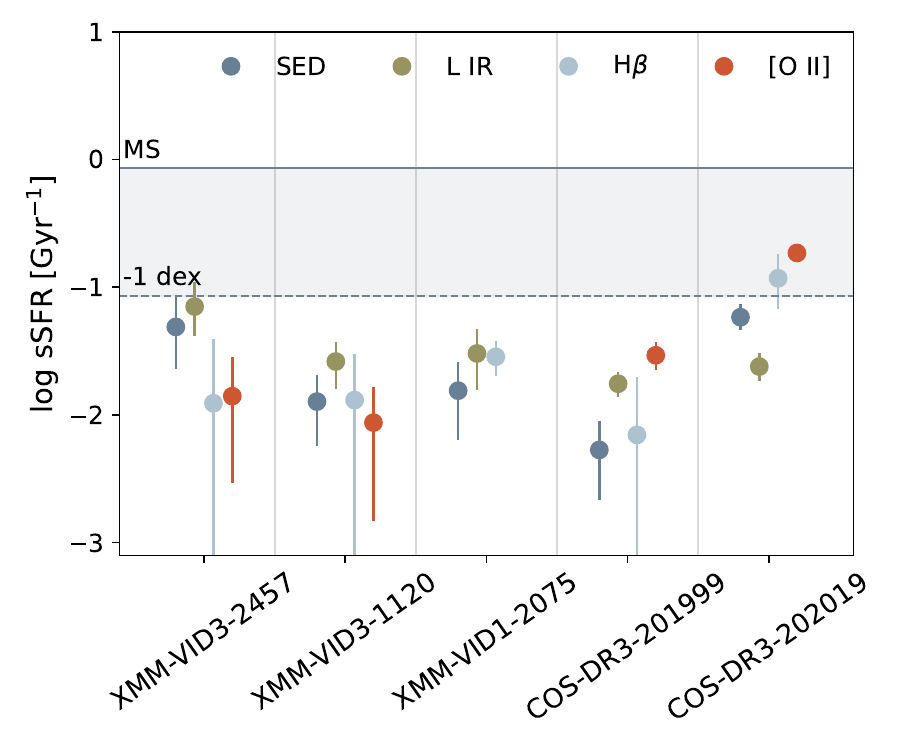}
\caption{Comparison of sSFR estimates from different sources: \texttt{CIGALE} SED fitting (dark blue), $L_{\rm IR}$ (olive)
, \Hbeta\ (light blue), \OIIshort\ (red). We adopt the \citet{Tomczak2016} relation to calculate the expected sSFR for main-sequence UMGs (\logM\ = 11.5), \mbox{$\mathrm{\log sSFR_{MS}/Gyr^{-1} = -0.07}$}. The region between the main sequence value and 1 dex below is highlighted in grey. 
\label{fig: SFR_all}}
\end{figure}

\subsection{Specific Star Formation Rate}
\label{subsec: sSFR}

The \texttt{CIGALE}-derived specific star formation rates (sSFRs) of our sample are all below \mbox{$\log (\mathrm{sSFR_{MS}/Gyr^{-1}}) < -1$} (Table\ \ref{tab:prop_all}), placing them well below the star-forming main sequence and confirming their quiescence.
In Figure\ \ref{fig: SFR_all}, we compare sSFRs of our UMGs with several important tracers.
We derive SFRs from the total dust luminosity ($L_{\rm dust}$) obtained from SED modeling, which quantifies the energy re-emitted by interstellar dust grains after disentangling and removing the contribution from AGN heating.
$L_{\mathrm{dust}}$ is then converted to SFR using the \citet{Kennicutt1998} calibration adjusted for a \citet{Chabrier2003} IMF. 
We note that $L_{\mathrm{dust}}$ may include a contribution from dust heated by evolved stellar populations, especially in quiescent or post-starburst systems.
The line-based SFRs are adopted from \citet{Forrest2020b} and are estimated from the H$\beta$ and \OIIshort\ luminosities using line fluxes measured with \texttt{slinefit} (\citealt{slinefit}). 
XMM-VID3-2457, the only source with an ALMA detection, shows slightly elevated sSFRs from the SED fitting and $L_{\mathrm{IR}}$ relative to those inferred from the nebular emission lines, likely indicating a modest amount of dust-obscured star formation consistent with the detected thermal dust emission detected by ALMA.
Nevertheless, the sSFR estimates remain consistent within $1\sigma$ under the threshold of \mbox{$\log(\mathrm{sSFR/Gyr^{-1}}) < -1$}, and therefore still satisfies our quiescent selection criteria.
The ALMA detection thus does not preclude its classification as quiescent, but rather indicates the presence of weak, dust-obscured star formation in an otherwise quiescent UMG.
 

The other two XMM-field UMGs, XMM-VID3-1120 and XMM-VID1-2075, show no ALMA detection. This lack of significant dust emission is consistent with their low sSFRs derived from other tracers, confirming that they are robustly quiescent systems with \mbox{$\mathrm{\log (sSFR/Gyr^{-1}) < -1.5}$} below the SFMS. 
Even though COS-DR3-201999 shows strong AGN signatures (\mbox{$\mathrm{frac_{AGN}} > 0.4$}; see Table~\ref{tab:prop_all}), its SFRs remain well below the star-forming main sequence, indicating that the galaxy is quenched and the AGN activity is dominated by radio emission (see Figure\ \ref{fig: SED_all} and Section\ \ref{subsec: AGN}).
In contrast, COS-DR3-202019 display moderately elevated sSFRs inferred from \Hbeta\ and \OIIshort\ relative to SED- or $L_{\mathrm{IR}}$-based values, approaching the $-1$ dex threshold.
As seen in the figure, the sSFR derived from \OIIshort\ is higher by over 0.5 dex compared to the SED fit.  
COS-DR3-202019 was first reported in \citep{Forrest2020b} with a large uncertainty on estimation of SFR with \mbox{82.4$^{+10.4}_{-58.2} M_{\odot}/$yr}.
However, its ALMA non-detection rules out dusty obscured star formation, and its radio emission is likely associated with AGN activity (see Section~\ref{subsec: AGN}). 
While the strong emission of \OIIshort\ and \Hbeta\ could be produced from unobscured star formation, it could also arise from the narrow-line region of an AGN.
With the AGN contribution inferred from SED and the non-negligible radio AGN emission, we therefore caution that the SFRs from the line fluxes for COS-DR3-202019 could be significantly overestimated.


\subsection{AGN Contribution}
\label{subsec: AGN}

We utilize the \texttt{SKIRTOR} module in \texttt{CIGALE} to estimate the AGN fraction, frac$_{\rm AGN}$, defined as \mbox{$L_{\rm dust, AGN}/$($L_{\rm dust, AGN}$ + $L_{\rm dust, galaxy}$)}.
The two UMGs in the COSMOS field, COS-DR3-202019 and COS-DR3-201999, exhibit substantial AGN contribution (\texttt{CIGALE}-derived \mbox{$\mathrm{frac_{AGN} > 0.45}$}), while the other three UMGs in the XMM-field show faint AGN contributions with \mbox{$\mathrm{frac_{AGN} <0.2}$}.
COS-DR3-202019 shows strong detections at both 1.4 and 3~GHz (S/N~$>3$), while COS-DR3-201999 is constrained by radio upper limits. 
With a spectral index of $\alpha = -0.8$ \citep{Marsan2017}, we derive \mbox{$L_{\mathrm{1.4\,GHz}}< 4.0 \times10^{24}$~W~Hz$^{-1}$} for COS-DR3-201999 and the luminosity of \mbox{$L_{\mathrm{1.4\,GHz}} = (6.2 \pm 1.2)\times10^{24}$~W~Hz$^{-1}$} for COS-DR3-202019 following the equation from \citet{Butler2018}. 
In addition, the radio-loudness parameter from \texttt{CIGALE} radio module, defined as \mbox{$R_{\mathrm{AGN}} = L_{\nu,\,5\,\mathrm{GHz}} / L_{\nu,\,2500\,\text{\AA}}$}, is \mbox{$19 \pm 10$} for COS-DR3-201999 and higher for COS-DR3-202019 (\mbox{$144 \pm 45$}), both exceeding the conventional threshold of $R_{\mathrm{AGN}}>10$ \citep{Yang2022} and identifying them as radio-loud AGN, indicative of the powerful AGN jets.  

To assess the role of AGN emission in the UV-to-radio SEDs for two COS-UMGs, we rerun the SED fitting excluding AGN module while while keeping all other model components and parameter ranges identical to those in Section\ \ref{sec: SED}.
As shown in Figure\ \ref{fig: SED_Compare_AGN}, the non-AGN models underpredict the observed radio flux, indicating a radio excess relative to stellar processes alone, whereas the inclusion of an AGN component reproduce fits more accurately.
It confirms that the radio emission in two COS-UMGs, particularly COS-DR3-202019 with robust radio detections, is dominated by an AGN rather than residual star formation.
In \citet{Forrest2020b}, \OIIIshort\ and \Hbeta\ emission lines were tentatively detected in the MOSFIRE $K$-band spectrum of COS-DR3-201999, while the H$\beta$ feature is highly uncertain due to the low signal-to-noise.
The MOSFIRE spectra of the remaining XMM-field targets are dominated by stellar absorption features, with no robust detections of nebular emission lines.  
The corresponding line ratios are listed together with other AGN indicators in Table~\ref{tab:AGN_classification}. 
The AGN identification for the two COSMOS UMGs is mainly driven by their radio and SED properties. Their weak spectral line constraints are consistent with a low-excitation, jet-dominated AGN mode, as observed in low-excitation radio galaxies \citep[LERGS,][]{Butler2018}.

\begin{deluxetable}{lcccc}
\tabletypesize{\scriptsize}  
\centering
\caption{AGN classification. 
\label{tab:AGN_classification}}
\tablehead{
\colhead{UMG} &  \colhead{$f_{\rm AGN}$} & \colhead{$R_{\rm AGN}$} & \colhead{L$_{\rm 1.4GHz}$} & \colhead{Line Ratio} \\
\colhead{} & \colhead{} & \colhead{} & \colhead{\text{(10$^{24}$ W Hz$^{-1}$)}} & \colhead{$f_{\rm [OIII]5007}/f_{H\beta}$\tablenotemark{a}}  
}
\startdata
X–2457 & $0.16 \pm 0.11$ & -- & -- & $7.0 \pm 12.3$ \\
X–1120 & $0.14 \pm 0.09$ & -- &  -- &$-5.2\pm 6.3$  \\
X–2075 & $0.11 \pm 0.05$ & -- &  -- & $0.6 \pm 0.8$  \\
C–201999 &  $0.50 \pm 0.10$ & $>10$ & $<4$ & $8.7 \pm 14.9$ \\
C–202019 & $0.47 \pm 0.10$ & $>100$ & $6.2\pm1.2$ & $1.0 \pm 0.4$  \\
\enddata
\tablenotetext{a}{$f_{\rm [OIII]\lambda5007}/f_{H\beta} > 6$, consistent with AGNs at these large stellar masses.}
\end{deluxetable}

\section{Dust Properties}
\label{sec: Dust}
 
\subsection{SED-derived Dust Mass}
\label{subsec: DustMass} 

We estimate dust masses using the \citet{Draine2014} dust emission models in \texttt{CIGALE}.  
This model accounts for both diffuse interstellar dust heated by the general stellar population and dust in star-forming regions exposed to a range of radiation fields ($U_{\mathrm{min}}$-$U_{\mathrm{max}}$). The parameters $\gamma$ and $q_{\mathrm{PAH}}$ control the fraction of warm dust and small aromatic grains, allowing a consistent derivation of total dust mass.
We present dust mass ($M_{\rm dust,SED}$) and dust luminosities ($L_{\rm dust}$) for five UMGs in Table\ \ref{tab:prop_all}, where $L_{\rm dust}$ is adopted to estimate SFR$_{\rm IR}$ in Section\ \ref{subsec: sSFR}).
The ALMA-detected source shows the highest dust mass with $\log(M_{\rm dust}/M_\odot)=8.43^{+0.15}_{-0.22}$, whereas the remaining galaxies have significantly lower values with $\log(M_{\rm dust}/M_\odot)<8$, listed in Table\ \ref{tab:prop_all}. 



\subsection{Modified Blackbody Dust Mass}
\label{subsec: MBB}

Another efficient method for estimating dust mass from the single-band FIR detection is to apply a modified blackbody (MBB) fit. We determine dust masses according to the following equation:
\[
M_{\rm dust} = \frac{D_L^2 S_{\nu_o}}{(1+z)\,\kappa_{\nu_r}}
\left[ B_{\nu_r}(T_d) - B_{\nu_r}(T_{\mathrm{CMB}}(z)) \right]^{-1},
\]
where $S_{\nu_o}$ is the flux density at the ALMA observed frequency (345 GHz for band 7) with \mbox{$\nu_o = \nu_r (1+z)^{-1}$}. 
We adopt \mbox{$\kappa_{\nu_r}=\kappa_0 (\lambda_0/\lambda_{\rm rest})^{\beta}$}, assuming a dust mass opacity coefficient $\kappa_0$ of $0.0431~\mathrm{m^{2}~kg^{-1}}$ at $\lambda_0 = 850 \mu m$ \citep{LiDraine2001}, and a dust emissivity index $\beta$ of 1.8 \citep{Scoville2014}.
\mbox{$T_{\mathrm{CMB}}(z) = 2.725 \times (1 + z)$} is the CMB temperature at redshift $z$, which affects the derived dust mass by $\sim$ 5\% ($\approx 0.02$~dex). 
The \citet{Draine2014} dust emission fits the effective dust temperature of \mbox{$T_d\sim22-33$~K} for our UMGs. We adopt a representative value of \mbox{$T_d = 25$~K} for the following analysis, which is consistent with the assumptions used for the literature MQGs compared in Section\ \ref{subsec: Dust-poor}.
For the four UMGs without ALMA detections, the values of $S_{\nu_0}$ correspond to $3\sigma$ upper limits, and the derived $M_{\rm dust}$ is also considered as upper limits.
The estimated dust masses are shown in Table\ \ref{tab:prop_all}. 

\begin{figure}
\centering
\includegraphics[width = 0.45\textwidth]{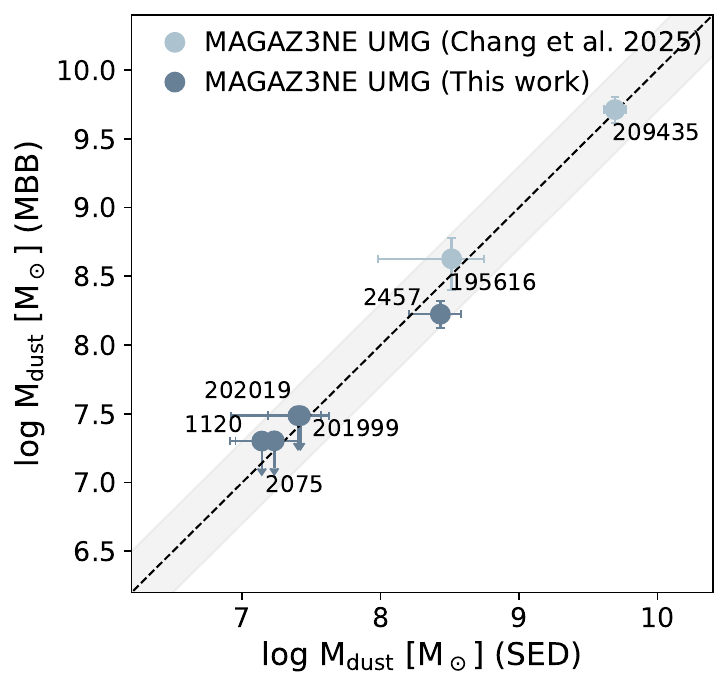}
\caption{Comparison between dust masses from \texttt{CIGALE} SED fitting and single-temperature modified blackbody fits for five MAGAZ3NE UMGs. The black dashed line is the 1:1 relation, and the grey band marks \(\pm0.3\)~dex (a factor of $\sim 2$) around unity. Source IDs are annotated.
\label{fig: Mdust_SED_MBB}}
\end{figure}

We compare MBB- and SED-derived dust masses in Fig.~\ref{fig: Mdust_SED_MBB}, which are broadly consistent. The robust MBB detection (XMM-VID3-2457) lies essentially on the 1:1 relation within 0.3 dex, indicating no obvious systematic offset in the SED-based dust mass scale. For the remaining galaxies, the MBB constraints are upper limits derived from ALMA non-detections, and all fall at or slightly above the SED estimates. 
Overall, we find no significant discrepancy between the two methods, with differences within $\lesssim0.3$~dex. 
The lower SED-derived dust masses for the two COSMOS targets, COS-DR3-201999 and COS-DR3-202019, are encompassed by the MBB-derived upper limits. 
In the following section, we utilize the MBB-derived dust masses and corresponding upper limits to discuss the dust content of our sample, including both detected and non-detected UMGs.

\begin{figure*}
\centering
\includegraphics[width = 0.48\textwidth]{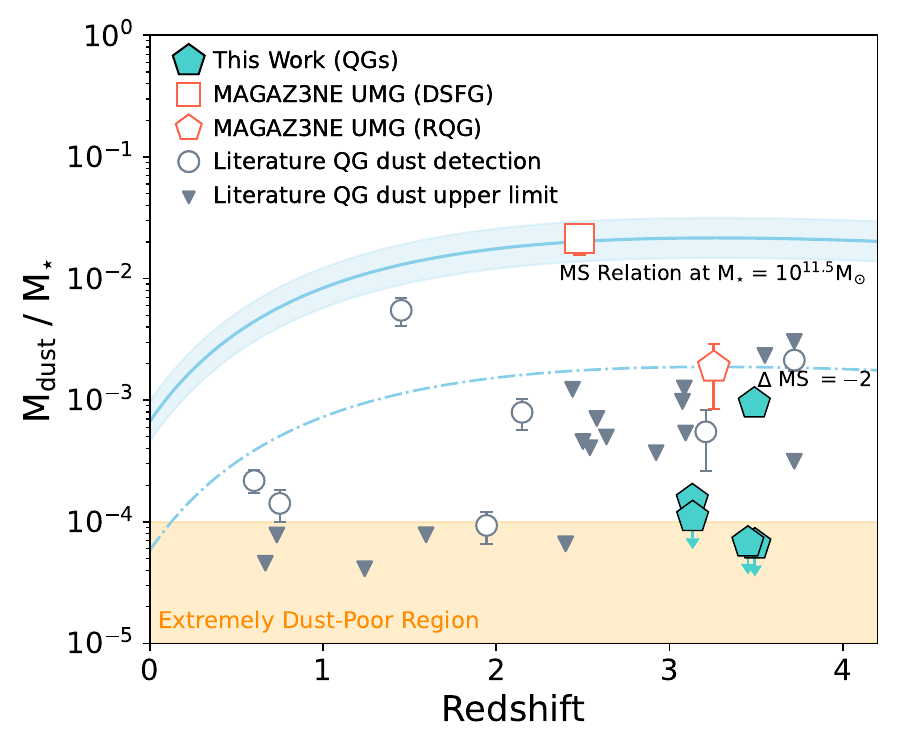}
\includegraphics[width = 0.48\textwidth]{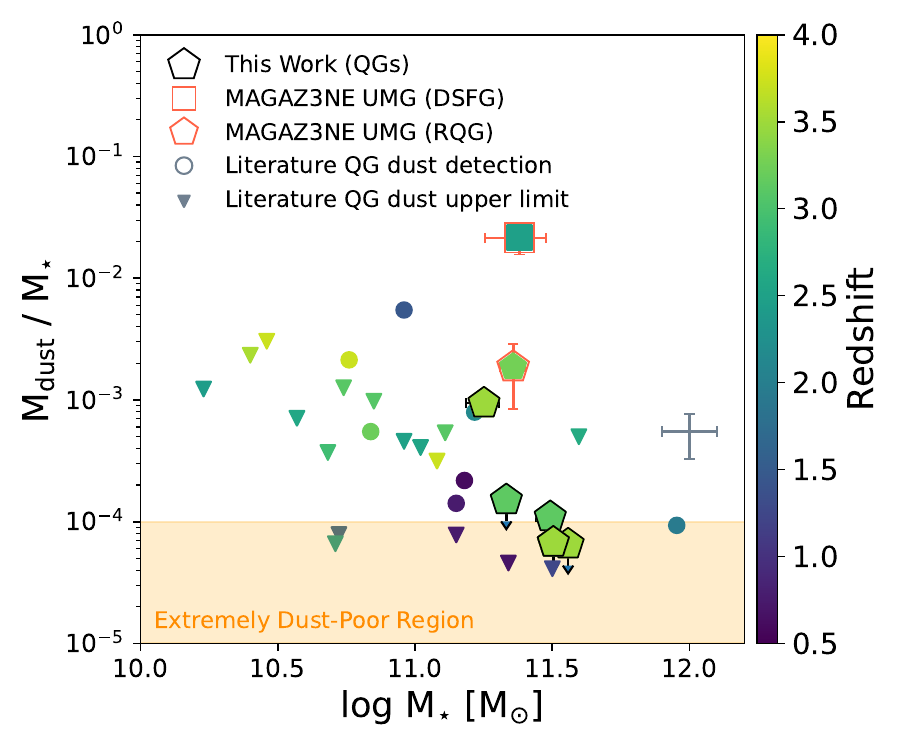}
\caption{
Dust mass fraction ($M_{\rm dust}/M_{\star}$) of massive quiescent galaxies as a function of redshift (left) and stellar mass (right).
Five UMGs analyzed in this work are shown as filled cyan pentagons in the left panel (black outlined pentagons in the right), together with two MAGAZ3NE UMGs with ALMA Band 7 detections (red square for one dusty star-forming galaxy and red pentagon for one recent quenching galaxy; \citealt{Chang2025arXiv}).  
Spectroscopically confirmed massive quiescent galaxies from the literature \citep{Gobat2022, Suzuki2022, Umehata2025, Spilker2025arXiv,Siegel2025} are shown as open circles when detected in dust emission and as inverted triangles when only upper limits are available.
In the right panel, the literature galaxies are color-coded by redshift, and the gray cross indicates the median uncertainties in stellar mass and the dust-to-stellar mass ratio.
The blue shaded region in the left panel represents the main-sequence (MS) scaling relation at $M_\star = 10^{11-12}~M_\odot$ from \citet{Tacconi2018}, and the blue dash-dotted line shows galaxies offset by $\Delta{\mathrm{MS}}=-2$.
The orange-shaded region shows the extremely dust-poor regime with \mbox{$M_{\rm dust}/M_{\star} = 10^{-5} - 10^{-4}$}.  
\label{fig: frac_Mdust}}
\end{figure*}

\subsection{Extremely Dust-poor MQGs}
\label{subsec: Dust-poor}

In Figure\ \ref{fig: frac_Mdust} (left), we show the dust-to-stellar mass ratio ($M_{\rm dust}/M_{\star}$) for five UMGs, derived from MBB dust mass estimation and stellar mass fitted from the SED fitting using \texttt{CIGALE}.  
All five UMGs occupy the extreme low end of the $M_{\rm dust}/M_{\star}$ distribution, with 2 dex below the star-forming scaling relation \citep{Tacconi2018}.  
Among them, XMM-VID3-2457 is the only source detected in dust emission, exhibiting the highest $M_{\rm dust}/M_{\star}$ ratio of \mbox{$(0.94 \pm 0.17)\times10^{-3}$}.
XMM-VID3-1120 and XMM-VID1-2075 fall within the dust-poor orange-shaded region, with $M_{\mathrm{dust}}/M_\star < 10^{-4}$, indicating an almost complete depletion of the cold interstellar medium and consistent with their classification as quiescent systems. 
The remaining two UMGs in the COSMOS field, COS-DR3-202019 and COS-DR3-201999, show the relatively low ratio of $M_{\mathrm{dust}}/M_\star$ near $10^{-4}$. 
We include literature measurements of spectroscopically confirmed massive quiescent galaxies at $0.5 < z < 4$, shown as detections with error bars and as upper limits for non-detections.
\citet{Whitaker2021, Gobat2022} confirmed and performed detailed SED fitting for the same lensed quiescent massive galaxies at $1.5 < z < 3.2$, finding $M_{\mathrm{dust}}/M_\star$ ratios down to $\sim10^{-4}$ for a subset of their sample, while others exhibit moderately higher values.
\citet{Spilker2025arXiv, Siegel2025} identified a small sample of dust-deficient massive quiescent galaxies with only upper limits on dust mass at $0.7 < z < 2.4$. 
In the higher-redshift range $2.5 < z < 4$, \citet{Suzuki2022}, \citet{Siegel2025}, and \citet{Umehata2025} reported quiescent systems with comparatively higher $M_{\mathrm{dust}}/M_\star$ ratios of $2\times10^{-4}-5\times10^{-3}$, suggesting that these galaxies are not entirely dust-poor.
An exception is ALMA.14 at $z=1.45$ \citep{Hayashi2018}, which exhibits a comparatively high dust-to-stellar mass ratio and may not represent a fully quiescent system \citep{Lee2024}.
Most other massive quiescent galaxies in the literature at $z\gtrsim3$ remain undetected in dust emission, with detections reported for only 2 of 9 sources and upper limits for the remaining 7 sources.

In Figure~\ref{fig: frac_Mdust} (right), we show the dust-to-stellar mass ratio ($M_{\mathrm{dust}}/M_\star$) as a function of stellar mass.
The UMGs analyzed in this work (black pentagons) lie at the lowest end of the distribution, with $M_{\mathrm{dust}}/M_\star < 10^{-4}$ for three sources (XMM-VID3-1120, XMM-VID1-2075, and COS-DR3-201999), placing them among the most dust-poor quiescent galaxies identified spectroscopically at $3 < z < 4$. 
This apparent dust deficiency at $z>3$ could plausibly reflect rapid dust depletion or removal processes operating in the most massive quiescent galaxies in the early universe. 
It could provide new observational constraints on dust consumption and quenching processes in the early universe.

\begin{figure*}
\centering
\includegraphics[width = 0.98\textwidth]{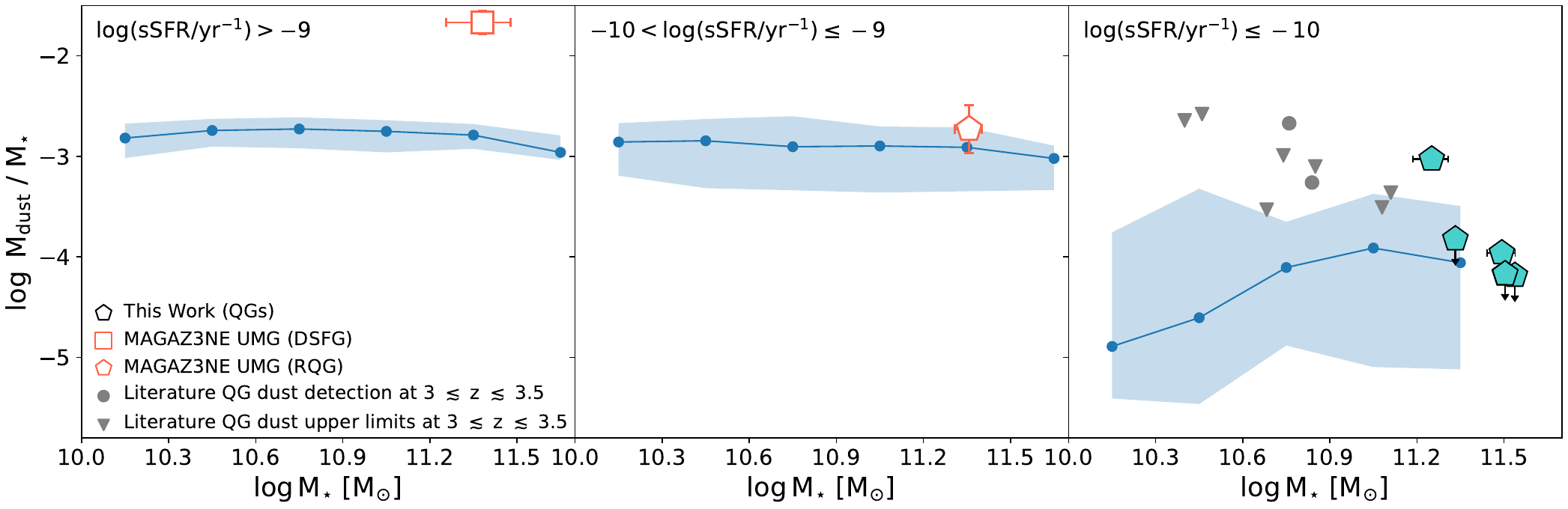} 
\caption{Dust-to-stellar mass ratio as a function of stellar mass for UMGs at
$3 \le z \le 3.5$, separated into three bins of specific star-formation rate:
$\log(\mathrm{sSFR}/\mathrm{yr}^{-1}) > -9$ (left),
$-10 < \log(\mathrm{sSFR}/\mathrm{yr}^{-1}) \le -9$ (middle), and
$\log(\mathrm{sSFR}/\mathrm{yr}^{-1}) \le -10$ (right).
Blue points and shaded regions indicate the median and the 16th-84th percentile range of $\log(M_{\mathrm{dust}}/M_\star)$ in each stellar-mass bin from the SIMBA simulation described in Section~\ref{subsec: Dust-poor}.
Grey symbols represent individual massive quiescent galaxies from the literature observed with ALMA at $3 \lesssim z \lesssim 3.5$ \citep{Gobat2022, Suzuki2022, Umehata2025}, with inverted triangles indicating upper limits, same as Figure\ \ref{fig: frac_Mdust}.
MAGAZ3NE UMGs in this work are shown as filled cyan pentagons with two UMGs (red open square for one dusty star-forming galaxy and red open pentagon for one recent quenching galaxy) in \citet{Chang2025arXiv}.
\label{fig: SIMBA}}
\end{figure*}

To place our results in theoretical models, we compare our measurements with predictions from the SIMBA cosmological simulation \citep{Dave2019} in Figure~\ref{fig: SIMBA}. 
We select galaxies from multiple snapshots spanning $ 3<z\lesssim3.5$ of the $(100\,h^{-1}\,\mathrm{Mpc})^3$ simulation box and restrict the sample to massive systems with $\log(M_\star/M_\odot) > 10$. The selected galaxies are divided into bins of specific star formation rate: $\log(\mathrm{sSFR}/\mathrm{yr}^{-1}) > -9$, $-10 < \log(\mathrm{sSFR}/\mathrm{yr}^{-1}) \le -9$, and $\log(\mathrm{sSFR}/\mathrm{yr}^{-1}) \le -10$, with the lowest sSFR bin corresponding to the quiescent population. 
SIMBA derived $M_{\rm dust}/M_\star$ ratio depends on sSFR, with a broad dispersion in the lowest-sSFR bin. 
Four non-detected quiescent UMGs in this work lie at the extreme high-mass end (\logM $>11.3$) of the simulated SIMBA quiescent population. 
While they show broad agreement with the predicted trends of $M_{\rm dust}/M_\star$ in the quiescent population, four UMGs occupy a region that remains sparsely sampled in current simulations at $3<z\lesssim3.5$, effectively extending the parameter space sampled by SIMBA.
In contrast, MQGs observed with ALMA continuum in the literature with similar redshifts \citep{Gobat2022, Suzuki2022, Umehata2025} exhibit elevated $M_{\rm dust}/M_\star$ ratios relative to the SIMBA median, even though the majority of these MQGs are non-detections.  
This suggest that they represent quiescent galaxies at an earlier stage of the quenching process, retaining modest dust reservoirs despite their low sSFRs.



The low dust detection fraction complicates gas-mass estimates based on dust emission.
Recent studies have reported unusually high gas-to-dust ratios, reaching values up to 1000 times in massive quiescent galaxies at $0.5 < z < 1.3$ \citep{Spilker2025arXiv}, suggesting that the conversion from dust mass to gas mass carries significant uncertainty.
The commonly adopted assumption of a constant gas-to-dust ratio of $\sim$100 may therefore be unreliable in the absence of direct constraints, even for high-redshift systems, making gas-mass estimation particularly challenging for dust-deficient galaxies such as the UMGs presented here.
Further constraints from direct molecular gas tracers, such as CO or [CI] line observations, are thus essential for accurately determining gas masses and verifying whether these galaxies are truly gas-poor or instead exhibit atypically high gas-to-dust ratios.

\section{Conclusions} 
\label{sec: conc}

In this work, we presented the ALMA observation with Band 7 dust continuum for five UMGs at $z>3$ in the MAGAZ3NE survey. 
All five UMGs are located within the quiescent region of the $UVJ$ diagram.
Only one galaxy, XMM-VID3-2457, exhibits a faint ALMA Band 7 continuum detection, while the remaining four are undetected down to a sensitivity of \mbox{$\sigma_{RMS} = 18.6$ $\rm \mu Jy$ $\rm beam^{-1}$}.  

By incorporating ALMA constraints into the \texttt{CIGALE} analysis, we derived reliable stellar, dust, and star-formation properties for five UMGs. Our results confirm that the UV-NIR-selected galaxies are not heavily obscured star-forming systems but lie more than one dex below the star-forming main sequence with \mbox{$\mathrm{\log (sSFR/Gyr^{-1}) < -1}$}, consistent with strongly suppressed star formation as the quiescent galaxies. We further find evidence for AGN activity in a subset of this quiescent sample, including two radio-loud galaxies, while the remaining systems show weak or no AGN signatures.
 
We estimate dust masses using both SED modeling and modified blackbody fitting, with consistent results between the two methods. 
The dust-to-stellar mass ratios of our UMGs are more than two dex below the scaling relation for the main-sequence galaxies.
The ALMA-detected UMG, XMM-VID3-2457, remains relatively dust-rich with $M_{\mathrm{dust}}/M_\star \sim 10^{-3}$.
Four UMGs fall in or near the extremely dust-poor, with $M_{\mathrm{dust}}/M_\star \lesssim 10^{-4}$, indicating severe dust depletion during the quenching phase. These values are comparable to or lower than those of spectroscopically confirmed massive quiescent galaxies reported in the literature, placing our sample among the extremely dust-poor systems known at $3 < z < 4$.  
We compare our results with the SIMBA simulation across different star-formation activity bins at $3 < z \lesssim 3.5$ and find that while our four quiescent UMGs show broad agreement with predicted trends, they occupy a high-mass region with \logM $>11.3$ that remains sparsely sampled.


This work reveals a significant dust deficiency in the most massive quiescent galaxies at $z > 3$, showing that the dust reservoir may have been depleted more rapidly during the quenching process than previously thought. 
With the deep ALMA constraints, we show that dust-poor UMGs can be robustly distinguished from heavily obscured star-forming systems, even at the highest stellar masses. 
Our findings could provide new observational constraints on rapid dust-removal processes during early quenching, where these processes remain poorly explored in simulations of the most massive galaxies.




\begin{acknowledgments}

The authors wish to recognize and acknowledge the very significant cultural role and reverence that the summit of Maunakea has always had within the indigenous Hawaiian community. We are most fortunate to have the opportunity to conduct observations from this mountain. 
Based on data products from observations made with ESO Telescopes at the La Silla Paranal Observatory under ESO programme ID 179.A-2005 and on data products produced by CALET and the Cambridge Astronomy Survey Unit on behalf of the UltraVISTA consortium.

GW gratefully acknowledges support from the National Science Foundation through grant AST-2347348. 
BF acknowledges support from JWST-GO-02913.001-A.
AM acknowledges support from the Yavin Family Fund.
This work is supported by grant number 80NSSC21K0630 issued through
the NASA Astrophysics Data Analysis Program (ADAP) and by the National Science Foundation through grant AST-2009442.  

\end{acknowledgments}

%
 
\software{astropy \citep{2013A&A...558A..33A,2018AJ....156..123A,2022ApJ...935..167A}, \texttt{CIGALE} \cite{Boquien2019, Yang2020, Yang2022}
          }


\appendix
\renewcommand{\thefigure}{A.\arabic{figure}}
\setcounter{figure}{0}

\renewcommand{\thetable}{A.\arabic{table}}
\setcounter{table}{0}

\begin{deluxetable}{lccc}
\tabletypesize{\footnotesize}   
\tablecaption{\texttt{CIGALE} modules and input parameters for the SED fitting. \label{tab:SED_input}} 
\tablehead{
\colhead{Module} & \colhead{Parameter} & \colhead{Name in \texttt{CIGALE}} & \colhead{Values}
}
\startdata
Delayed SFH
  & e-folding time of main population  & tau\_main & 100, 150, 300, 500, 1000, 3000 Myr\\
  & Age of main population& age\_main & 700, 1000, 1300, 1600 Myr\\
  & Age of the late burst & age\_burst & 50, 100, 150, 200, 250, 300 Myr \\
  & Mass fraction of the late burst population & f\_burst &  0.001, 0.005, 0.01, 0.05, 0.1, 0.3 \\
\hline
SSP & Initial mass function & imf & Chabrier (2003) \\
  & Stellar metallicity & metallicity  & 0.02 \\
\hline
Nebular emission
  & Ionization parameter & logU & $-4.0$, $-3.0$, $-2.0$,  \\
  & Gas metallicity & z\_gas  & 0.008, 0.014, 0.02, 0.03 \\
\hline
Dust attenuation 
  & Color excess of nebular emission lines & E\_BV\_lines  & 0, 0.04, 0.07, 0.1, 0.2, 0.3, 0.4, 0.5 \\
  & Amplitude of the UV bump & uv\_bump\_amplitude & 0.0, 1, 2, 3 \\
  & Slope of the attenuation curve & powerlaw\_slope &  $-2$, $-1.5$,  $-1$, $-0.5$, $0.0$ \\
\hline
Dust emission
  & Mass fraction of PAH & qpah & 1.77, 2.5, 3.19 \\
  & Minimum radiation field & umin & 0.1, 0.5, 1, 1.5, 2, 4, 6, 10 \\
  & Powerlaw slope &  alpha & 1.5, 2, 2.5 \\
  & Fraction illuminated from Umin to Umax &  gamma  & 0.01, 0.05, 0.1, 0.2 \\
\hline
AGN emission
  & Inclination & i & $30^{\circ}$, $70^{\circ}$ \\
  & AGN fraction & fracAGN& 0.1--0.9 (step 0.1) \\
  & Extinction in the polar direction & EBV & 0.03, 0.1, 0.2, 0.3 \\
\hline
Radio \tablenotemark{a} 
& The slope of synchrotron emission & alpha\_sf & 0.8 \\
  & The radio-loudness parameter & $R_{\rm AGN}$ & 0.1, 1, 10, 50, 100, 150, 200 \\
\enddata
\tablenotetext{a}{This module is only fitted for UMGs in the COSMOS field (COS-DR3-201999 and COS-DR3-202019).}
\end{deluxetable}

\begin{figure*}[!t]
\centering
\includegraphics[width = 0.21\textwidth]{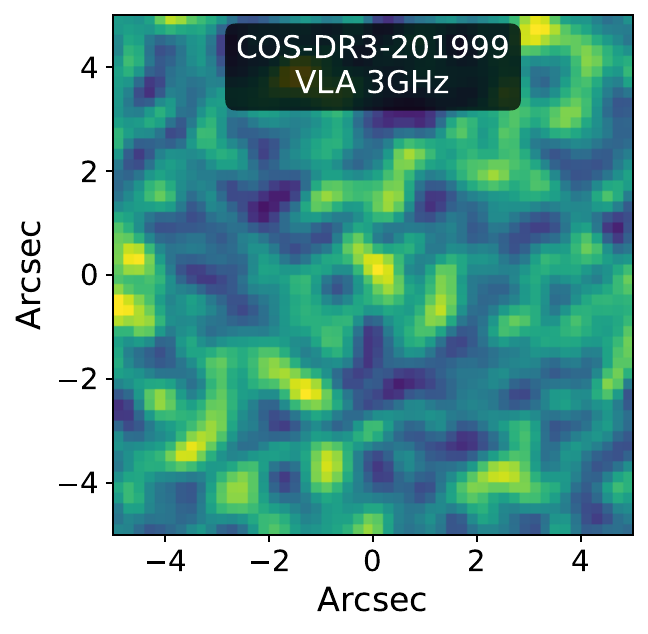}
\includegraphics[width = 0.21\textwidth]{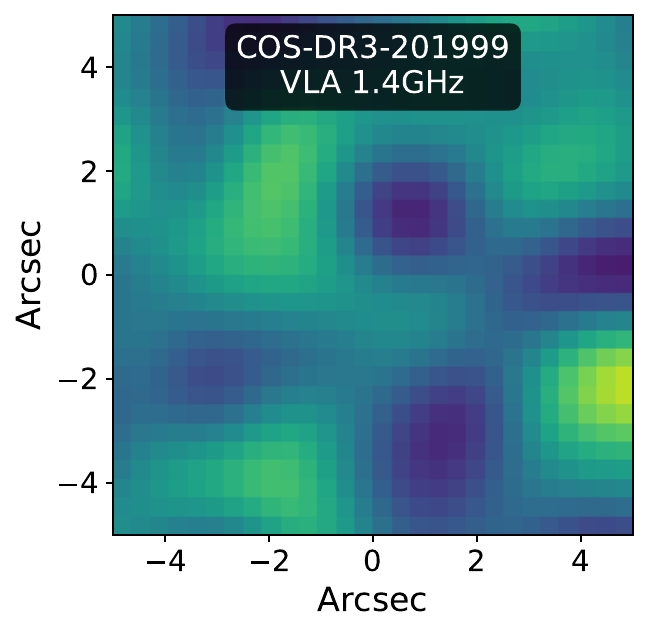}
\includegraphics[width = 0.55\textwidth]{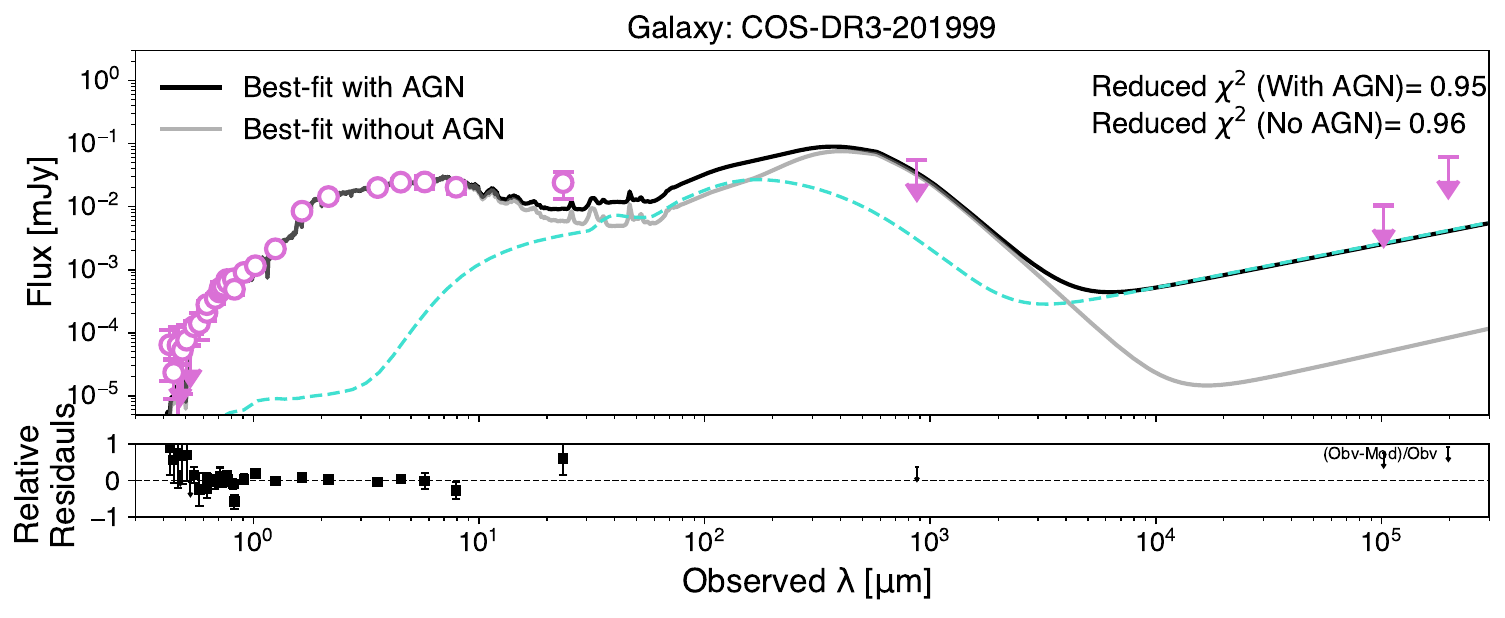}
\includegraphics[width = 0.21\textwidth]{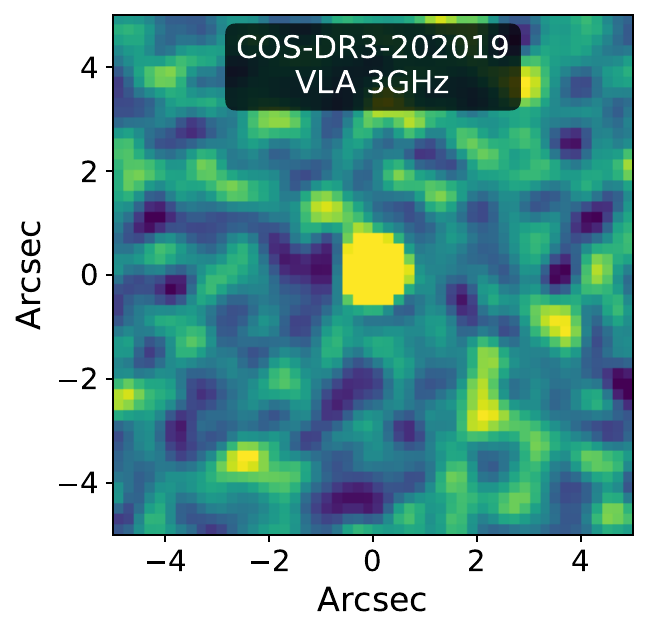}
\includegraphics[width = 0.21\textwidth]{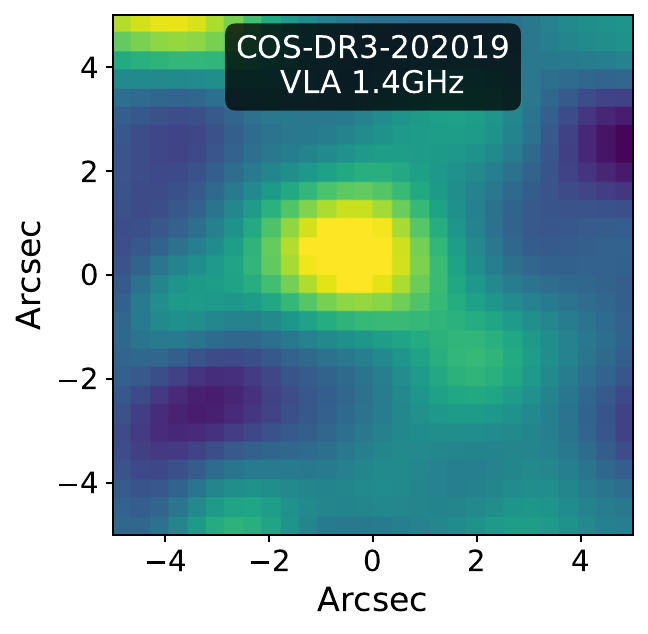}
\includegraphics[width = 0.55\textwidth]{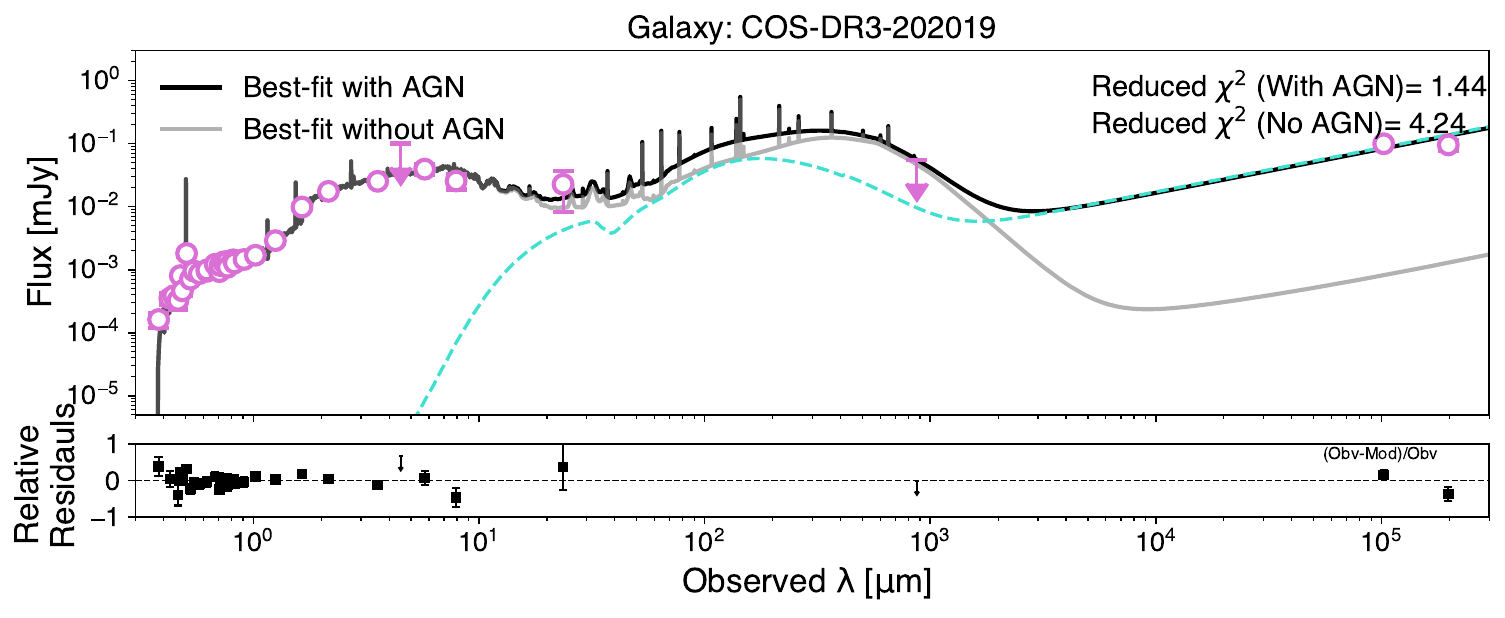} 
\caption{VLA radio images and UV-to-radio SED modeling for COS-DR3-201999 ($z_{\rm spec} = 3.131$) and COS-DR3-202019 ($z_{\rm spec} = 3.133$), shown in the same format as Figure\ \ref{fig: SED_all}. Black and gray curves represent the best-fit models with (black solid line) and without (gray solid line) an AGN component, respectively, with the AGN contribution shown by the cyan dashed curve.  
The reduced $\chi^{2}$ values for the fits with and without AGN are listed in the panel titles.
\label{fig: SED_Compare_AGN}}
\end{figure*}

\bibliography{main}{}

@ARTICLE{Kennicutt1998,
       author = {{Kennicutt}, Jr., Robert C.},
        title = "{Star Formation in Galaxies Along the Hubble Sequence}",
      journal = {\araa},
         year = 1998,
        month = jan,
       volume = {36},
        pages = {189-232},
          doi = {10.1146/annurev.astro.36.1.189},
archivePrefix = {arXiv},
       eprint = {astro-ph/9807187},
 primaryClass = {astro-ph},
       adsurl = {https://ui.adsabs.harvard.edu/abs/1998ARA&A..36..189K}
}

@software{slinefit,
author = {Corentin Schreiber},
title = {cschreib/slinefit},
version = {v2.3},
publisher = {GitHub},
year  = {2023},
url = {https://github.com/cschreib/slinefit}
}

@software{fastpp,
author = {Corentin Schreiber and Hugh Dickinson},
title = {cschreib/fastpp},
version = {v1.5.0},
publisher = {GitHub},
year  = {2023},
url = {https://github.com/cschreib/fastpp}
}

@ARTICLE{Weaver2022,
       author = {{Weaver}, J.~R. and {Kauffmann}, O.~B. and {Ilbert}, O. and {McCracken}, H.~J. and {Moneti}, A. and {Toft}, S. and {Brammer}, G. and {Shuntov}, M. and {Davidzon}, I. and {Hsieh}, B.~C. and {Laigle}, C. and {Anastasiou}, A. and {Jespersen}, C.~K. and {Vinther}, J. and {Capak}, P. and {Casey}, C.~M. and {McPartland}, C.~J.~R. and {Milvang-Jensen}, B. and {Mobasher}, B. and {Sanders}, D.~B. and {Zalesky}, L. and {Arnouts}, S. and {Aussel}, H. and {Dunlop}, J.~S. and {Faisst}, A. and {Franx}, M. and {Furtak}, L.~J. and {Fynbo}, J.~P.~U. and {Gould}, K.~M.~L. and {Greve}, T.~R. and {Gwyn}, S. and {Kartaltepe}, J.~S. and {Kashino}, D. and {Koekemoer}, A.~M. and {Kokorev}, V. and {Le F{\`e}vre}, O. and {Lilly}, S. and {Masters}, D. and {Magdis}, G. and {Mehta}, V. and {Peng}, Y. and {Riechers}, D.~A. and {Salvato}, M. and {Sawicki}, M. and {Scarlata}, C. and {Scoville}, N. and {Shirley}, R. and {Silverman}, J.~D. and {Sneppen}, A. and {Smolc̆i{\'c}}, V. and {Steinhardt}, C. and {Stern}, D. and {Tanaka}, M. and {Taniguchi}, Y. and {Teplitz}, H.~I. and {Vaccari}, M. and {Wang}, W. -H. and {Zamorani}, G.},
        title = "{COSMOS2020: A Panchromatic View of the Universe to z{\ensuremath{\sim}}10 from Two Complementary Catalogs}",
      journal = {\apjs},
         year = 2022,
        month = jan,
       volume = {258},
       number = {1},
          eid = {11},
        pages = {11},
          doi = {10.3847/1538-4365/ac3078},
archivePrefix = {arXiv},
       eprint = {2110.13923},
 primaryClass = {astro-ph.GA},
       adsurl = {https://ui.adsabs.harvard.edu/abs/2022ApJS..258...11W}
}

@ARTICLE{Gobat2018,
       author = {{Gobat}, R. and {Daddi}, E. and {Magdis}, G. and {Bournaud}, F. and {Sargent}, M. and {Martig}, M. and {Jin}, S. and {Finoguenov}, A. and {B{\'e}thermin}, M. and {Hwang}, H.~S. and {Renzini}, A. and {Wilson}, G.~W. and {Aretxaga}, I. and {Yun}, M. and {Strazzullo}, V. and {Valentino}, F.},
        title = "{The unexpectedly large dust and gas content of quiescent galaxies at z > 1.4}",
      journal = {Nature Astronomy},
         year = 2018,
        month = jan,
       volume = {2},
        pages = {239-246},
          doi = {10.1038/s41550-017-0352-5},
archivePrefix = {arXiv},
       eprint = {1703.02207},
 primaryClass = {astro-ph.GA},
       adsurl = {https://ui.adsabs.harvard.edu/abs/2018NatAs...2..239G}
}

@ARTICLE{Lee2024,
       author = {{Lee}, Minju M. and {Steidel}, Charles C. and {Brammer}, Gabriel and {F{\"o}rster-Schreiber}, Natascha and {Renzini}, Alvio and {Liu}, Daizhong and {Herrera-Camus}, Rodrigo and {Naab}, Thorsten and {Price}, Sedona H. and {{\"U}bler}, Hannah and {Arriagada-Neira}, Sebasti{\'a}n and {Magdis}, Georgios},
        title = "{High dust content of a quiescent galaxy at z   2 revealed by deep ALMA observation}",
      journal = {\mnras},
         year = 2024,
        month = feb,
       volume = {527},
       number = {4},
        pages = {9529-9547},
          doi = {10.1093/mnras/stad3718},
archivePrefix = {arXiv},
       eprint = {2311.00023},
 primaryClass = {astro-ph.GA},
       adsurl = {https://ui.adsabs.harvard.edu/abs/2024MNRAS.527.9529L}
}

@ARTICLE{Spilker2025arXiv,
       author = {{Spilker}, Justin S. and {Whitaker}, Katherine E. and {Narayanan}, Desika and {Bezanson}, Rachel and {Bodansky}, Sarah and {D'Onofrio}, Vincenzo R. and {Feldmann}, Robert and {Goulding}, Andy D. and {Greene}, Jenny E. and {Kriek}, Mariska and {Luo}, Yuanze and {Setton}, David J. and {Suess}, Katherine A. and {van der Wel}, Arjen and {Verrico}, Margaret E. and {Williams}, Christina C. and {Woodrum}, Charity and {Wu}, Po-Feng},
        title = "{Unusually High Gas-to-Dust Ratios Observed in High-Redshift Quiescent Galaxies}",
      journal = {arXiv e-prints},
         year = 2025,
        month = jul,
          eid = {arXiv:2507.16914},
        pages = {arXiv:2507.16914},
          doi = {10.48550/arXiv.2507.16914},
archivePrefix = {arXiv},
       eprint = {2507.16914},
 primaryClass = {astro-ph.GA},
       adsurl = {https://ui.adsabs.harvard.edu/abs/2025arXiv250716914S}
}

@ARTICLE{Siegel2025,
       author = {{Siegel}, Jared C. and {Setton}, David J. and {Greene}, Jenny E. and {Suess}, Katherine A. and {Whitaker}, Katherine E. and {Bezanson}, Rachel and {Leja}, Joel and {Furtak}, Lukas J. and {Cutler}, Sam E. and {de Graaff}, Anna and {Feldmann}, Robert and {Khullar}, Gourav and {Labbe}, Ivo and {Marchesini}, Danilo and {Miller}, Tim B. and {Nanayakkara}, Themiya and {Pan}, Richard and {Price}, Sedona H. and {Treiber}, Helena P. and {van Dokkum}, Pieter and {Wang}, Bingjie and {Weaver}, John R.},
        title = "{UNCOVER: Significant Reddening in Cosmic Noon Quiescent Galaxies}",
      journal = {\apj},
         year = 2025,
        month = may,
       volume = {985},
       number = {1},
          eid = {125},
        pages = {125},
          doi = {10.3847/1538-4357/adc7b7},
archivePrefix = {arXiv},
       eprint = {2409.11457},
 primaryClass = {astro-ph.GA},
       adsurl = {https://ui.adsabs.harvard.edu/abs/2025ApJ...985..125S}
}

@ARTICLE{Suzuki2022,
       author = {{Suzuki}, Tomoko L. and {Glazebrook}, Karl and {Schreiber}, Corentin and {Kodama}, Tadayuki and {Kacprzak}, Glenn G. and {Leiton}, Roger and {Nanayakkara}, Themiya and {Oesch}, Pascal A. and {Papovich}, Casey and {Spitler}, Lee and {Straatman}, Caroline M.~S. and {Tran}, Kim-Vy and {Wang}, Tao},
        title = "{Low Star Formation Activity and Low Gas Content of Quiescent Galaxies at z = 3.5-4.0 Constrained with ALMA}",
      journal = {\apj},
         year = 2022,
        month = sep,
       volume = {936},
       number = {1},
          eid = {61},
        pages = {61},
          doi = {10.3847/1538-4357/ac7ce3},
archivePrefix = {arXiv},
       eprint = {2206.14238},
 primaryClass = {astro-ph.GA},
       adsurl = {https://ui.adsabs.harvard.edu/abs/2022ApJ...936...61S}
}

@ARTICLE{Umehata2025,
       author = {{Umehata}, Hideki and {Kubo}, Mariko and {Nakanishi}, Kouichiro},
        title = "{ADF22-WEB: Detection of a Molecular Gas Reservoir in a Massive Quiescent Galaxy Located in a z {\ensuremath{\approx}} 3 Protocluster Core}",
      journal = {\apjl},
         year = 2025,
        month = may,
       volume = {985},
       number = {1},
          eid = {L8},
        pages = {L8},
          doi = {10.3847/2041-8213/add1d4},
archivePrefix = {arXiv},
       eprint = {2502.06538},
 primaryClass = {astro-ph.GA},
       adsurl = {https://ui.adsabs.harvard.edu/abs/2025ApJ...985L...8U}
}

@ARTICLE{Gobat2022,
       author = {{Gobat}, R. and {D'Eugenio}, C. and {Liu}, D. and {Caminha}, G.~B. and {Daddi}, E. and {Bl{\'a}nquez}, D.},
        title = "{The uncertain interstellar medium of high-redshift quiescent galaxies: Impact of methodology}",
      journal = {\aap},
         year = 2022,
        month = dec,
       volume = {668},
          eid = {L4},
        pages = {L4},
          doi = {10.1051/0004-6361/202244995},
archivePrefix = {arXiv},
       eprint = {2211.14131},
 primaryClass = {astro-ph.GA},
       adsurl = {https://ui.adsabs.harvard.edu/abs/2022A&A...668L...4G}
}

@ARTICLE{Hayashi2018,
       author = {{Hayashi}, Masao and {Tadaki}, Ken-ichi and {Kodama}, Tadayuki and {Kohno}, Kotaro and {Yamaguchi}, Yuki and {Hatsukade}, Bunyo and {Koyama}, Yusei and {Shimakawa}, Rhythm and {Tamura}, Yoichi and {Suzuki}, Tomoko L.},
        title = "{Molecular Gas Reservoirs in Cluster Galaxies at z = 1.46}",
      journal = {\apj},
         year = 2018,
        month = apr,
       volume = {856},
       number = {2},
          eid = {118},
        pages = {118},
          doi = {10.3847/1538-4357/aab3e7},
archivePrefix = {arXiv},
       eprint = {1803.00298},
 primaryClass = {astro-ph.GA},
       adsurl = {https://ui.adsabs.harvard.edu/abs/2018ApJ...856..118H}
}

@ARTICLE{Kubo2022,
       author = {{Kubo}, Mariko and {Umehata}, Hideki and {Matsuda}, Yuichi and {Kajisawa}, Masaru and {Steidel}, Charles C. and {Yamada}, Toru and {Tanaka}, Ichi and {Hatsukade}, Bunyo and {Tamura}, Yoichi and {Nakanishi}, Kouichiro and {Kohno}, Kotaro and {Lee}, Kianhong and {Matsuda}, Keiichi and {Ao}, Yiping and {Nagao}, Tohru and {Yun}, Min S.},
        title = "{An AGN with an Ionized Gas Outflow in a Massive Quiescent Galaxy in a Protocluster at z = 3.09}",
      journal = {\apj},
         year = 2022,
        month = aug,
       volume = {935},
       number = {2},
          eid = {89},
        pages = {89},
          doi = {10.3847/1538-4357/ac7f2d},
archivePrefix = {arXiv},
       eprint = {2207.03628},
 primaryClass = {astro-ph.GA},
       adsurl = {https://ui.adsabs.harvard.edu/abs/2022ApJ...935...89K}
}

@ARTICLE{Chang2025arXiv,
       author = {{Chang}, Wenjun and {Wilson}, Gillian and {Forrest}, Ben and {McConachie}, Ian and {Webb}, Tracy and {Noble}, Allison G. and {Muzzin}, Adam and {Cooper}, Michael C. and {Marchesini}, Danilo and {Canalizo}, Gabriela and {Battisti}, A.~J. and {Le Bail}, Aur{\'e}lien and {Gomez}, Percy L. and {Urbano Stawinski}, Stephanie M. and {Wisz}, Marie E.},
        title = "{MAGAZ3NE: Far-IR and Radio Insights into the Nature and Properties of Ultramassive Galaxies at $zrsim3$}",
      journal = {arXiv e-prints},
         year = 2025,
        month = aug,
          eid = {arXiv:2508.08460},
        pages = {arXiv:2508.08460},
          doi = {10.48550/arXiv.2508.08460},
archivePrefix = {arXiv},
       eprint = {2508.08460},
 primaryClass = {astro-ph.GA},
       adsurl = {https://ui.adsabs.harvard.edu/abs/2025arXiv250808460C}
}

@ARTICLE{Tomczak2016,
       author = {{Tomczak}, Adam R. and {Quadri}, Ryan F. and {Tran}, Kim-Vy H. and {Labb{\'e}}, Ivo and {Straatman}, Caroline M.~S. and {Papovich}, Casey and {Glazebrook}, Karl and {Allen}, Rebecca and {Brammer}, Gabreil B. and {Cowley}, Michael and {Dickinson}, Mark and {Elbaz}, David and {Inami}, Hanae and {Kacprzak}, Glenn G. and {Morrison}, Glenn E. and {Nanayakkara}, Themiya and {Persson}, S. Eric and {Rees}, Glen A. and {Salmon}, Brett and {Schreiber}, Corentin and {Spitler}, Lee R. and {Whitaker}, Katherine E.},
        title = "{The SFR-M* Relation and Empirical Star-Formation Histories from ZFOURGE* at 0.5 < z < 4}",
      journal = {\apj},
         year = 2016,
        month = feb,
       volume = {817},
       number = {2},
          eid = {118},
        pages = {118},
          doi = {10.3847/0004-637X/817/2/118},
archivePrefix = {arXiv},
       eprint = {1510.06072},
 primaryClass = {astro-ph.GA},
       adsurl = {https://ui.adsabs.harvard.edu/abs/2016ApJ...817..118T}
}

@ARTICLE{LiDraine2001,
       author = {{Li}, Aigen and {Draine}, B.~T.},
        title = "{Infrared Emission from Interstellar Dust. II. The Diffuse Interstellar Medium}",
      journal = {\apj},
         year = 2001,
        month = jun,
       volume = {554},
       number = {2},
        pages = {778-802},
          doi = {10.1086/323147},
archivePrefix = {arXiv},
       eprint = {astro-ph/0011319},
 primaryClass = {astro-ph},
       adsurl = {https://ui.adsabs.harvard.edu/abs/2001ApJ...554..778L}
}

@ARTICLE{Bugiani2025,
       author = {{Bugiani}, Letizia and {Belli}, Sirio and {Park}, Minjung and {Davies}, Rebecca L. and {Mendel}, J. Trevor and {Johnson}, Benjamin D. and {Khoram}, Amir H. and {Benton}, Chlo{\"e} and {Cimatti}, Andrea and {Conroy}, Charlie and {Emami}, Razieh and {Leja}, Joel and {Li}, Yijia and {Maheson}, Gabriel and {Mathews}, Elijah P. and {Naidu}, Rohan P. and {Nelson}, Erica J. and {Tacchella}, Sandro and {Terrazas}, Bryan A. and {Weinberger}, Rainer},
        title = "{Active Galactic Nucleus Feedback in Quiescent Galaxies at Cosmic Noon Traced by Ionized Gas Emission}",
      journal = {\apj},
         year = 2025,
        month = mar,
       volume = {981},
       number = {1},
          eid = {25},
        pages = {25},
          doi = {10.3847/1538-4357/adaeaf},
archivePrefix = {arXiv},
       eprint = {2406.08547},
 primaryClass = {astro-ph.GA},
       adsurl = {https://ui.adsabs.harvard.edu/abs/2025ApJ...981...25B}
}

@ARTICLE{Nanayakkara2025,
       author = {{Nanayakkara}, Themiya and {Glazebrook}, Karl and {Schreiber}, Corentin and {Chittenden}, Harry and {Brammer}, Gabriel and {Esdaile}, James and {Jacobs}, Colin and {Kacprzak}, Glenn G. and {Kawinwanichakij}, Lalitwadee and {Kimmig}, Lucas C. and {Labbe}, Ivo and {Lagos}, Claudia and {Marchesini}, Danilo and {Mart{\`\i}nez-Mar{\`\i}n}, M. and {Marsan}, Z. Cemile and {Oesch}, Pascal A. and {Papovich}, Casey and {Remus}, Rhea-Silvia and {Tran}, Kim-Vy H.},
        title = "{The Formation Histories of Massive and Quiescent Galaxies in the 3 < z < 4.5 Universe}",
      journal = {\apj},
         year = 2025,
        month = mar,
       volume = {981},
       number = {1},
          eid = {78},
        pages = {78},
          doi = {10.3847/1538-4357/ada6ac},
archivePrefix = {arXiv},
       eprint = {2410.02076},
 primaryClass = {astro-ph.GA},
       adsurl = {https://ui.adsabs.harvard.edu/abs/2025ApJ...981...78N}
}

@ARTICLE{Whitaker2021,
       author = {{Whitaker}, Katherine E. and {Williams}, Christina C. and {Mowla}, Lamiya and {Spilker}, Justin S. and {Toft}, Sune and {Narayanan}, Desika and {Pope}, Alexandra and {Magdis}, Georgios E. and {van Dokkum}, Pieter G. and {Akhshik}, Mohammad and {Bezanson}, Rachel and {Brammer}, Gabriel B. and {Leja}, Joel and {Man}, Allison and {Nelson}, Erica J. and {Richard}, Johan and {Pacifici}, Camilla and {Sharon}, Keren and {Valentino}, Francesco},
        title = "{Quenching of star formation from a lack of inflowing gas to galaxies}",
      journal = {\nat},
         year = 2021,
        month = sep,
       volume = {597},
       number = {7877},
        pages = {485-488},
          doi = {10.1038/s41586-021-03806-7},
archivePrefix = {arXiv},
       eprint = {2109.10384},
 primaryClass = {astro-ph.GA},
       adsurl = {https://ui.adsabs.harvard.edu/abs/2021Natur.597..485W}
}

@ARTICLE{Bezanson2019,
       author = {{Bezanson}, Rachel and {Spilker}, Justin and {Williams}, Christina C. and {Whitaker}, Katherine E. and {Narayanan}, Desika and {Weiner}, Benjamin and {Franx}, Marijn},
        title = "{Extremely Low Molecular Gas Content in a Compact, Quiescent Galaxy at z = 1.522}",
      journal = {\apjl},
         year = 2019,
        month = mar,
       volume = {873},
       number = {2},
          eid = {L19},
        pages = {L19},
          doi = {10.3847/2041-8213/ab0c9c},
archivePrefix = {arXiv},
       eprint = {1902.09564},
 primaryClass = {astro-ph.GA},
       adsurl = {https://ui.adsabs.harvard.edu/abs/2019ApJ...873L..19B}
}

@ARTICLE{Carnall2023,
       author = {{Carnall}, A.~C. and {McLeod}, D.~J. and {McLure}, R.~J. and {Dunlop}, J.~S. and {Begley}, R. and {Cullen}, F. and {Donnan}, C.~T. and {Hamadouche}, M.~L. and {Jewell}, S.~M. and {Jones}, E.~W. and {Pollock}, C.~L. and {Wild}, V.},
        title = "{A surprising abundance of massive quiescent galaxies at 3 < z < 5 in the first data from JWST CEERS}",
      journal = {\mnras},
         year = 2023,
        month = apr,
       volume = {520},
       number = {3},
        pages = {3974-3985},
          doi = {10.1093/mnras/stad369},
archivePrefix = {arXiv},
       eprint = {2208.00986},
 primaryClass = {astro-ph.GA},
       adsurl = {https://ui.adsabs.harvard.edu/abs/2023MNRAS.520.3974C}
}

@ARTICLE{Butler2018,
       author = {{Butler}, Andrew and {Huynh}, Minh and {Delvecchio}, Ivan and {Kapi{\'n}ska}, Anna and {Ciliegi}, Paolo and {Jurlin}, Nika and {Delhaize}, Jacinta and {Smol{\v{c}}i{\'c}}, Vernesa and {Desai}, Shantanu and {Fotopoulou}, Sotiria and {Lidman}, Chris and {Pierre}, Marguerite and {Plionis}, Manolis},
        title = "{The XXL Survey. XXXI. Classification and host galaxy properties of 2.1 GHz ATCA XXL-S radio sources}",
      journal = {\aap},
         year = 2018,
        month = nov,
       volume = {620},
          eid = {A16},
        pages = {A16},
          doi = {10.1051/0004-6361/201732379},
archivePrefix = {arXiv},
       eprint = {1804.05983},
 primaryClass = {astro-ph.GA},
       adsurl = {https://ui.adsabs.harvard.edu/abs/2018A&A...620A..16B}
}

@ARTICLE{Oke1983,
       author = {{Oke}, J.~B. and {Gunn}, J.~E.},
        title = "{Secondary standard stars for absolute spectrophotometry.}",
      journal = {\apj},
         year = 1983,
        month = mar,
       volume = {266},
        pages = {713-717},
          doi = {10.1086/160817},
       adsurl = {https://ui.adsabs.harvard.edu/abs/1983ApJ...266..713O}
}

@ARTICLE{Stawinski2024,
       author = {{Urbano Stawinski}, Stephanie M. and {Cooper}, M.~C. and {Forrest}, Ben and {Muzzin}, Adam and {Marchesini}, Danilo and {Wilson}, Gillian and {Gomez}, Percy and {McConachie}, Ian and {Marsan}, Z. Cemile and {Annuziatella}, Marianna and {Chang}, Wenjun},
        title = "{Spectroscopic Confirmation of an Ultra-Massive Galaxy in a Protocluster at z {\ensuremath{\sim}} 4.9}",
      journal = {The Open Journal of Astrophysics},
         year = 2024,
        month = jun,
       volume = {7},
          eid = {46},
        pages = {46},
          doi = {10.33232/001c.120087},
archivePrefix = {arXiv},
       eprint = {2404.16036},
 primaryClass = {astro-ph.GA},
       adsurl = {https://ui.adsabs.harvard.edu/abs/2024OJAp....7E..46U}
}

@ARTICLE{McConachie2025,
       author = {{McConachie}, Ian and {Wilson}, Gillian and {Forrest}, Ben and {Marsan}, Z. Cemile and {Muzzin}, Adam and {Cooper}, M.~C. and {Annunziatella}, Marianna and {Marchesini}, Danilo and {Gomez}, Percy and {Chang}, Wenjun and {Urbano Stawinski}, Stephanie M. and {McDonald}, Michael and {Webb}, Tracy and {Noble}, Allison and {Lemaux}, Brian C. and {Shah}, Ekta A. and {Staab}, Priti and {Lubin}, Lori M. and {Gal}, Roy R.},
        title = "{MAGAZ3NE: Evidence for Galactic Conformity in z {\ensuremath{\gtrsim}} 3 Protoclusters}",
      journal = {\apj},
         year = 2025,
        month = jan,
       volume = {978},
       number = {1},
          eid = {17},
        pages = {17},
          doi = {10.3847/1538-4357/ad8f36},
archivePrefix = {arXiv},
       eprint = {2411.14641},
 primaryClass = {astro-ph.GA},
       adsurl = {https://ui.adsabs.harvard.edu/abs/2025ApJ...978...17M}
}

@ARTICLE{Forrest2024b,
       author = {{Forrest}, Ben and {Cooper}, M.~C. and {Muzzin}, Adam and {Wilson}, Gillian and {Marchesini}, Danilo and {McConachie}, Ian and {Gomez}, Percy and {Annunziatella}, Marianna and {Marsan}, Z. Cemile and {Braspenning}, Joey and {Chang}, Wenjun and {de Lucia}, Gabriella and {Fontanot}, Fabio and {Hirschmann}, Michaela and {Nelson}, Dylan and {Pillepich}, Annalisa and {Schaye}, Joop and {Urbano Stawinski}, Stephanie M. and {Stefanon}, Mauro and {Xie}, Lizhi},
        title = "{MAGAZ3NE: Massive, Extremely Dusty Galaxies at z {\ensuremath{\sim}} 2 Lead to Photometric Overestimation of Number Densities of the Most Massive Galaxies at 3 < z < 4}",
      journal = {\apj},
         year = 2024,
        month = dec,
       volume = {977},
       number = {1},
          eid = {51},
        pages = {51},
          doi = {10.3847/1538-4357/ad8b1c},
archivePrefix = {arXiv},
       eprint = {2404.19018},
 primaryClass = {astro-ph.GA},
       adsurl = {https://ui.adsabs.harvard.edu/abs/2024ApJ...977...51F}
}

@ARTICLE{Mauduit2012,
       author = {{Mauduit}, J. -C. and {Lacy}, M. and {Farrah}, D. and {Surace}, J.~A. and {Jarvis}, M. and {Oliver}, S. and {Maraston}, C. and {Vaccari}, M. and {Marchetti}, L. and {Zeimann}, G. and {Gonz{\'a}les-Solares}, E.~A. and {Pforr}, J. and {Petric}, A.~O. and {Henriques}, B. and {Thomas}, P.~A. and {Afonso}, J. and {Rettura}, A. and {Wilson}, G. and {Falder}, J.~T. and {Geach}, J.~E. and {Huynh}, M. and {Norris}, R.~P. and {Seymour}, N. and {Richards}, G.~T. and {Stanford}, S.~A. and {Alexander}, D.~M. and {Becker}, R.~H. and {Best}, P.~N. and {Bizzocchi}, L. and {Bonfield}, D. and {Castro}, N. and {Cava}, A. and {Chapman}, S. and {Christopher}, N. and {Clements}, D.~L. and {Covone}, G. and {Dubois}, N. and {Dunlop}, J.~S. and {Dyke}, E. and {Edge}, A. and {Ferguson}, H.~C. and {Foucaud}, S. and {Franceschini}, A. and {Gal}, R.~R. and {Grant}, J.~K. and {Grossi}, M. and {Hatziminaoglou}, E. and {Hickey}, S. and {Hodge}, J.~A. and {Huang}, J. -S. and {Ivison}, R.~J. and {Kim}, M. and {LeFevre}, O. and {Lehnert}, M. and {Lonsdale}, C.~J. and {Lubin}, L.~M. and {McLure}, R.~J. and {Messias}, H. and {Mart{\'\i}nez-Sansigre}, A. and {Mortier}, A.~M.~J. and {Nielsen}, D.~M. and {Ouchi}, M. and {Parish}, G. and {Perez-Fournon}, I. and {Pierre}, M. and {Rawlings}, S. and {Readhead}, A. and {Ridgway}, S.~E. and {Rigopoulou}, D. and {Romer}, A.~K. and {Rosebloom}, I.~G. and {Rottgering}, H.~J.~A. and {Rowan-Robinson}, M. and {Sajina}, A. and {Simpson}, C.~J. and {Smail}, I. and {Squires}, G.~K. and {Stevens}, J.~A. and {Taylor}, R. and {Trichas}, M. and {Urrutia}, T. and {van Kampen}, E. and {Verma}, A. and {Xu}, C.~K.},
        title = "{Addendum: The Spitzer Extragalactic Representative Volume Survey (SERVS): Survey Definition and Goals (<A href=``http:///cgi-bin/resolve?id=doi: 10.1086/666945''>PASP, 124, 714, [2012]</A>)}",
      journal = {\pasp},
         year = 2012,
        month = oct,
       volume = {124},
       number = {920},
        pages = {1135},
          doi = {10.1086/668290},
       adsurl = {https://ui.adsabs.harvard.edu/abs/2012PASP..124.1135M}
}

@ARTICLE{Schreiber2018a_J&H_FAST,
       author = {{Schreiber}, C. and {Labb{\'e}}, I. and {Glazebrook}, K. and {Bekiaris}, G. and {Papovich}, C. and {Costa}, T. and {Elbaz}, D. and {Kacprzak}, G.~G. and {Nanayakkara}, T. and {Oesch}, P. and {Pannella}, M. and {Spitler}, L. and {Straatman}, C. and {Tran}, K. -V. and {Wang}, T.},
        title = "{Jekyll \& Hyde: quiescence and extreme obscuration in a pair of massive galaxies 1.5 Gyr after the Big Bang}",
      journal = {\aap},
         year = 2018,
        month = mar,
       volume = {611},
          eid = {A22},
        pages = {A22},
          doi = {10.1051/0004-6361/201731917},
archivePrefix = {arXiv},
       eprint = {1709.03505},
 primaryClass = {astro-ph.GA},
       adsurl = {https://ui.adsabs.harvard.edu/abs/2018A&A...611A..22S}
}

@ARTICLE{Schreiber2018b_UVJ,
       author = {{Schreiber}, C. and {Glazebrook}, K. and {Nanayakkara}, T. and {Kacprzak}, G.~G. and {Labb{\'e}}, I. and {Oesch}, P. and {Yuan}, T. and {Tran}, K. -V. and {Papovich}, C. and {Spitler}, L. and {Straatman}, C.},
        title = "{Near infrared spectroscopy and star-formation histories of 3 {\ensuremath{\leq}} z {\ensuremath{\leq}} 4 quiescent galaxies}",
      journal = {\aap},
         year = 2018,
        month = oct,
       volume = {618},
          eid = {A85},
        pages = {A85},
          doi = {10.1051/0004-6361/201833070},
archivePrefix = {arXiv},
       eprint = {1807.02523},
 primaryClass = {astro-ph.GA},
       adsurl = {https://ui.adsabs.harvard.edu/abs/2018A&A...618A..85S}
}

@ARTICLE{2011Whitaker,
       author = {{Whitaker}, Katherine E. and {Labb{\'e}}, Ivo and {van Dokkum}, Pieter G. and {Brammer}, Gabriel and {Kriek}, Mariska and {Marchesini}, Danilo and {Quadri}, Ryan F. and {Franx}, Marijn and {Muzzin}, Adam and {Williams}, Rik J. and {Bezanson}, Rachel and {Illingworth}, Garth D. and {Lee}, Kyoung-Soo and {Lundgren}, Britt and {Nelson}, Erica J. and {Rudnick}, Gregory and {Tal}, Tomer and {Wake}, David A.},
        title = "{The NEWFIRM Medium-band Survey: Photometric Catalogs, Redshifts, and the Bimodal Color Distribution of Galaxies out to z \raisebox{-0.5ex}\textasciitilde 3}",
      journal = {\apj},
         year = 2011,
        month = jul,
       volume = {735},
       number = {2},
          eid = {86},
        pages = {86},
          doi = {10.1088/0004-637X/735/2/86},
archivePrefix = {arXiv},
       eprint = {1105.4609},
 primaryClass = {astro-ph.CO},
       adsurl = {https://ui.adsabs.harvard.edu/abs/2011ApJ...735...86W}
}

@ARTICLE{Jarvis2013,
       author = {{Jarvis}, Matt J. and {Bonfield}, D.~G. and {Bruce}, V.~A. and {Geach}, J.~E. and {McAlpine}, K. and {McLure}, R.~J. and {Gonz{\'a}lez-Solares}, E. and {Irwin}, M. and {Lewis}, J. and {Yoldas}, A. Kupcu and {Andreon}, S. and {Cross}, N.~J.~G. and {Emerson}, J.~P. and {Dalton}, G. and {Dunlop}, J.~S. and {Hodgkin}, S.~T. and {Le}, F{\`e}vre O. and {Karouzos}, M. and {Meisenheimer}, K. and {Oliver}, S. and {Rawlings}, S. and {Simpson}, C. and {Smail}, I. and {Smith}, D.~J.~B. and {Sullivan}, M. and {Sutherland}, W. and {White}, S.~V. and {Zwart}, J.~T.~L.},
        title = "{The VISTA Deep Extragalactic Observations (VIDEO) survey}",
      journal = {\mnras},
         year = 2013,
        month = jan,
       volume = {428},
       number = {2},
        pages = {1281-1295},
          doi = {10.1093/mnras/sts118},
archivePrefix = {arXiv},
       eprint = {1206.4263},
 primaryClass = {astro-ph.CO},
       adsurl = {https://ui.adsabs.harvard.edu/abs/2013MNRAS.428.1281J}
}

@ARTICLE{McConachie2022,
       author = {{McConachie}, Ian and {Wilson}, Gillian and {Forrest}, Ben and {Marsan}, Z. Cemile and {Muzzin}, Adam and {Cooper}, M.~C. and {Annunziatella}, Marianna and {Marchesini}, Danilo and {Chan}, Jeffrey C.~C. and {Gomez}, Percy and {Abdullah}, Mohamed H. and {Saracco}, Paolo and {Nantais}, Julie},
        title = "{Spectroscopic Confirmation of a Protocluster at z = 3.37 with a High Fraction of Quiescent Galaxies}",
      journal = {\apj},
         year = 2022,
        month = feb,
       volume = {926},
       number = {1},
          eid = {37},
        pages = {37},
          doi = {10.3847/1538-4357/ac2b9f},
archivePrefix = {arXiv},
       eprint = {2109.07696},
 primaryClass = {astro-ph.GA},
       adsurl = {https://ui.adsabs.harvard.edu/abs/2022ApJ...926...37M}
}

@ARTICLE{Saracco2020,
       author = {{Saracco}, Paolo and {Marchesini}, Danilo and {La Barbera}, Francesco and {Gargiulo}, Adriana and {Annunziatella}, Marianna and {Forrest}, Ben and {Lange Vagle}, Daniel J. and {Marsan}, Z. Cemile and {Muzzin}, Adam and {Stefanon}, Mauro and {Wilson}, Gillian},
        title = "{The Rapid Buildup of Massive Early-type Galaxies: Supersolar Metallicity, High Velocity Dispersion, and Young Age for an Early-type Galaxy at z = 3.35}",
      journal = {\apj},
         year = 2020,
        month = dec,
       volume = {905},
       number = {1},
          eid = {40},
        pages = {40},
          doi = {10.3847/1538-4357/abc7c4},
archivePrefix = {arXiv},
       eprint = {2011.04657},
 primaryClass = {astro-ph.GA},
       adsurl = {https://ui.adsabs.harvard.edu/abs/2020ApJ...905...40S}
}

@ARTICLE{Tacconi2018,
       author = {{Tacconi}, L.~J. and {Genzel}, R. and {Saintonge}, A. and {Combes}, F. and {Garc{\'\i}a-Burillo}, S. and {Neri}, R. and {Bolatto}, A. and {Contini}, T. and {F{\"o}rster Schreiber}, N.~M. and {Lilly}, S. and {Lutz}, D. and {Wuyts}, S. and {Accurso}, G. and {Boissier}, J. and {Boone}, F. and {Bouch{\'e}}, N. and {Bournaud}, F. and {Burkert}, A. and {Carollo}, M. and {Cooper}, M. and {Cox}, P. and {Feruglio}, C. and {Freundlich}, J. and {Herrera-Camus}, R. and {Juneau}, S. and {Lippa}, M. and {Naab}, T. and {Renzini}, A. and {Salome}, P. and {Sternberg}, A. and {Tadaki}, K. and {{\"U}bler}, H. and {Walter}, F. and {Weiner}, B. and {Weiss}, A.},
        title = "{PHIBSS: Unified Scaling Relations of Gas Depletion Time and Molecular Gas Fractions}",
      journal = {\apj},
         year = 2018,
        month = feb,
       volume = {853},
       number = {2},
          eid = {179},
        pages = {179},
          doi = {10.3847/1538-4357/aaa4b4},
archivePrefix = {arXiv},
       eprint = {1702.01140},
 primaryClass = {astro-ph.GA},
       adsurl = {https://ui.adsabs.harvard.edu/abs/2018ApJ...853..179T}
}

@ARTICLE{Hwang2021,
       author = {{Hwang}, Yu-Hsuan and {Wang}, Wei-Hao and {Chang}, Yu-Yen and {Lim}, Chen-Fatt and {Chen}, Chian-Chou and {Gao}, Zhen-Kai and {Dunlop}, James S. and {Gao}, Yu and {Ho}, Luis C. and {Hwang}, Ho Seong and {Koprowski}, Maciej and {Micha{\l}owski}, Micha{\l} J. and {Peng}, Ying-jie and {Shim}, Hyunjin and {Simpson}, James M. and {Toba}, Yoshiki},
        title = "{Revisiting the Color-Color Selection: Submillimeter and AGN Properties of NUV-r-J Selected Quiescent Galaxies}",
      journal = {\apj},
         year = 2021,
        month = may,
       volume = {913},
       number = {1},
          eid = {6},
        pages = {6},
          doi = {10.3847/1538-4357/abf11a},
archivePrefix = {arXiv},
       eprint = {2103.14336},
 primaryClass = {astro-ph.GA},
       adsurl = {https://ui.adsabs.harvard.edu/abs/2021ApJ...913....6H}
}

@ARTICLE{Scoville2007,
       author = {{Scoville}, N. and {Aussel}, H. and {Brusa}, M. and {Capak}, P. and {Carollo}, C.~M. and {Elvis}, M. and {Giavalisco}, M. and {Guzzo}, L. and {Hasinger}, G. and {Impey}, C. and {Kneib}, J. -P. and {LeFevre}, O. and {Lilly}, S.~J. and {Mobasher}, B. and {Renzini}, A. and {Rich}, R.~M. and {Sanders}, D.~B. and {Schinnerer}, E. and {Schminovich}, D. and {Shopbell}, P. and {Taniguchi}, Y. and {Tyson}, N.~D.},
        title = "{The Cosmic Evolution Survey (COSMOS): Overview}",
      journal = {\apjs},
         year = 2007,
        month = sep,
       volume = {172},
       number = {1},
        pages = {1-8},
          doi = {10.1086/516585},
archivePrefix = {arXiv},
       eprint = {astro-ph/0612305},
 primaryClass = {astro-ph},
       adsurl = {https://ui.adsabs.harvard.edu/abs/2007ApJS..172....1S}
}

@ARTICLE{Belli2014,
       author = {{Belli}, Sirio and {Newman}, Andrew B. and {Ellis}, Richard S. and {Konidaris}, Nick P.},
        title = "{MOSFIRE Absorption Line Spectroscopy of z > 2 Quiescent Galaxies: Probing a Period of Rapid Size Growth}",
      journal = {\apjl},
         year = 2014,
        month = jun,
       volume = {788},
       number = {2},
          eid = {L29},
        pages = {L29},
          doi = {10.1088/2041-8205/788/2/L29},
archivePrefix = {arXiv},
       eprint = {1404.4872},
 primaryClass = {astro-ph.GA},
       adsurl = {https://ui.adsabs.harvard.edu/abs/2014ApJ...788L..29B}
}

@ARTICLE{Glazebrook2017,
       author = {{Glazebrook}, Karl and {Schreiber}, Corentin and {Labb{\'e}}, Ivo and {Nanayakkara}, Themiya and {Kacprzak}, Glenn G. and {Oesch}, Pascal A. and {Papovich}, Casey and {Spitler}, Lee R. and {Straatman}, Caroline M.~S. and {Tran}, Kim-Vy H. and {Yuan}, Tiantian},
        title = "{A massive, quiescent galaxy at a redshift of 3.717}",
      journal = {\nat},
         year = 2017,
        month = apr,
       volume = {544},
       number = {7648},
        pages = {71-74},
          doi = {10.1038/nature21680},
archivePrefix = {arXiv},
       eprint = {1702.01751},
 primaryClass = {astro-ph.GA},
       adsurl = {https://ui.adsabs.harvard.edu/abs/2017Natur.544...71G}
}

@ARTICLE{Lacy2021,
       author = {{Lacy}, M. and {Surace}, J.~A. and {Farrah}, D. and {Nyland}, K. and {Afonso}, J. and {Brandt}, W.~N. and {Clements}, D.~L. and {Lagos}, C.~D.~P. and {Maraston}, C. and {Pforr}, J. and {Sajina}, A. and {Sako}, M. and {Vaccari}, M. and {Wilson}, G. and {Ballantyne}, D.~R. and {Barkhouse}, W.~A. and {Brunner}, R. and {Cane}, R. and {Clarke}, T.~E. and {Cooper}, M. and {Cooray}, A. and {Covone}, G. and {D'Andrea}, C. and {Evrard}, A.~E. and {Ferguson}, H.~C. and {Frieman}, J. and {Gonzalez-Perez}, V. and {Gupta}, R. and {Hatziminaoglou}, E. and {Huang}, J. and {Jagannathan}, P. and {Jarvis}, M.~J. and {Jones}, K.~M. and {Kimball}, A. and {Lidman}, C. and {Lubin}, L. and {Marchetti}, L. and {Martini}, P. and {McMahon}, R.~G. and {Mei}, S. and {Messias}, H. and {Murphy}, E.~J. and {Newman}, J.~A. and {Nichol}, R. and {Norris}, R.~P. and {Oliver}, S. and {Perez-Fournon}, I. and {Peters}, W.~M. and {Pierre}, M. and {Polisensky}, E. and {Richards}, G.~T. and {Ridgway}, S.~E. and {R{\"o}ttgering}, H.~J.~A. and {Seymour}, N. and {Shirley}, R. and {Somerville}, R. and {Strauss}, M.~A. and {Suntzeff}, N. and {Thorman}, P.~A. and {van Kampen}, E. and {Verma}, A. and {Wechsler}, R. and {Wood-Vasey}, W.~M.},
        title = "{A Spitzer survey of Deep Drilling Fields to be targeted by the Vera C. Rubin Observatory Legacy Survey of Space and Time}",
      journal = {\mnras},
         year = 2021,
        month = feb,
       volume = {501},
       number = {1},
        pages = {892-910},
          doi = {10.1093/mnras/staa3714},
archivePrefix = {arXiv},
       eprint = {2011.15030},
 primaryClass = {astro-ph.GA},
       adsurl = {https://ui.adsabs.harvard.edu/abs/2021MNRAS.501..892L}
}

@ARTICLE{Dave2019,
       author = {{Dav{\'e}}, Romeel and {Angl{\'e}s-Alc{\'a}zar}, Daniel and {Narayanan}, Desika and {Li}, Qi and {Rafieferantsoa}, Mika H. and {Appleby}, Sarah},
        title = "{SIMBA: Cosmological simulations with black hole growth and feedback}",
      journal = {\mnras},
         year = 2019,
        month = jun,
       volume = {486},
       number = {2},
        pages = {2827-2849},
          doi = {10.1093/mnras/stz937},
archivePrefix = {arXiv},
       eprint = {1901.10203},
 primaryClass = {astro-ph.GA},
       adsurl = {https://ui.adsabs.harvard.edu/abs/2019MNRAS.486.2827D}
}

@ARTICLE{Jin2024,
       author = {{Jin}, Shuowen and {Sillassen}, Nikolaj B. and {Magdis}, Georgios E. and {Brinch}, Malte and {Shuntov}, Marko and {Brammer}, Gabriel and {Gobat}, Raphael and {Valentino}, Francesco and {Carnall}, Adam C. and {Lee}, Minju and {Vijayan}, Aswin P. and {Gillman}, Steven and {Kokorev}, Vasily and {Le Bail}, Aur{\'e}lien and {Greve}, Thomas R. and {Gullberg}, Bitten and {Gould}, Katriona M.~L. and {Toft}, Sune},
        title = "{Cosmic Vine: A z = 3.44 large-scale structure hosting massive quiescent galaxies}",
      journal = {\aap},
         year = 2024,
        month = mar,
       volume = {683},
          eid = {L4},
        pages = {L4},
          doi = {10.1051/0004-6361/202348540},
archivePrefix = {arXiv},
       eprint = {2311.04867},
 primaryClass = {astro-ph.GA},
       adsurl = {https://ui.adsabs.harvard.edu/abs/2024A&A...683L...4J}
}

@ARTICLE{Graaff2025,
       author = {{de Graaff}, Anna and {Setton}, David J. and {Brammer}, Gabriel and {Cutler}, Sam and {Suess}, Katherine A. and {Labb{\'e}}, Ivo and {Leja}, Joel and {Weibel}, Andrea and {Maseda}, Michael V. and {Whitaker}, Katherine E. and {Bezanson}, Rachel and {Boogaard}, Leindert A. and {Cleri}, Nikko J. and {De Lucia}, Gabriella and {Franx}, Marijn and {Greene}, Jenny E. and {Hirschmann}, Michaela and {Matthee}, Jorryt and {McConachie}, Ian and {Naidu}, Rohan P. and {Oesch}, Pascal A. and {Price}, Sedona H. and {Rix}, Hans-Walter and {Valentino}, Francesco and {Wang}, Bingjie and {Williams}, Christina C.},
        title = "{Efficient formation of a massive quiescent galaxy at redshift 4.9}",
      journal = {Nature Astronomy},
         year = 2025,
        month = feb,
       volume = {9},
        pages = {280-292},
          doi = {10.1038/s41550-024-02424-3},
archivePrefix = {arXiv},
       eprint = {2404.05683},
 primaryClass = {astro-ph.GA},
       adsurl = {https://ui.adsabs.harvard.edu/abs/2025NatAs...9..280D}
}

@ARTICLE{Scoville2014,
       author = {{Scoville}, N. and {Aussel}, H. and {Sheth}, K. and {Scott}, K.~S. and {Sanders}, D. and {Ivison}, R. and {Pope}, A. and {Capak}, P. and {Vanden Bout}, P. and {Manohar}, S. and {Kartaltepe}, J. and {Robertson}, B. and {Lilly}, S.},
        title = "{The Evolution of Interstellar Medium Mass Probed by Dust Emission: ALMA Observations at z = 0.3-2}",
      journal = {\apj},
         year = 2014,
        month = mar,
       volume = {783},
       number = {2},
          eid = {84},
        pages = {84},
          doi = {10.1088/0004-637X/783/2/84},
archivePrefix = {arXiv},
       eprint = {1401.2987},
 primaryClass = {astro-ph.GA},
       adsurl = {https://ui.adsabs.harvard.edu/abs/2014ApJ...783...84S}
}

@ARTICLE{Marsan2017,
       author = {{Marsan}, Z. Cemile and {Marchesini}, Danilo and {Brammer}, Gabriel B. and {Geier}, Stefan and {Kado-Fong}, Erin and {Labb{\'e}}, Ivo and {Muzzin}, Adam and {Stefanon}, Mauro},
        title = "{A Spectroscopic Follow-up Program of Very Massive Galaxies at 3 < z < 4: Confirmation of Spectroscopic Redshifts, and a High Fraction of Powerful AGNs}",
      journal = {\apj},
         year = 2017,
        month = jun,
       volume = {842},
       number = {1},
          eid = {21},
        pages = {21},
          doi = {10.3847/1538-4357/aa7206},
archivePrefix = {arXiv},
       eprint = {1606.05350},
 primaryClass = {astro-ph.GA},
       adsurl = {https://ui.adsabs.harvard.edu/abs/2017ApJ...842...21M}
}

@ARTICLE{Yang2020,
       author = {{Yang}, G. and {Boquien}, M. and {Buat}, V. and {Burgarella}, D. and {Ciesla}, L. and {Duras}, F. and {Stalevski}, M. and {Brandt}, W.~N. and {Papovich}, C.},
        title = "{X-CIGALE: Fitting AGN/galaxy SEDs from X-ray to infrared}",
      journal = {\mnras},
         year = 2020,
        month = jan,
       volume = {491},
       number = {1},
        pages = {740-757},
          doi = {10.1093/mnras/stz3001},
archivePrefix = {arXiv},
       eprint = {2001.08263},
 primaryClass = {astro-ph.GA},
       adsurl = {https://ui.adsabs.harvard.edu/abs/2020MNRAS.491..740Y}
}

@ARTICLE{Forrest2022,
       author = {{Forrest}, Ben and {Wilson}, Gillian and {Muzzin}, Adam and {Marchesini}, Danilo and {Cooper}, M.~C. and {Marsan}, Z. Cemile and {Annunziatella}, Marianna and {McConachie}, Ian and {Zaidi}, Kumail and {Gomez}, Percy and {Urbano Stawinski}, Stephanie M. and {Chang}, Wenjun and {de Lucia}, Gabriella and {La Barbera}, Francesco and {Lubin}, Lori and {Nantais}, Julie and {Pe{\~n}a}, Theodore and {Saracco}, Paolo and {Surace}, Jason and {Stefanon}, Mauro},
        title = "{MAGAZ3NE: High Stellar Velocity Dispersions for Ultramassive Quiescent Galaxies at z {\ensuremath{\gtrsim}} 3}",
      journal = {\apj},
         year = 2022,
        month = oct,
       volume = {938},
       number = {2},
          eid = {109},
        pages = {109},
          doi = {10.3847/1538-4357/ac8747},
archivePrefix = {arXiv},
       eprint = {2208.04329},
 primaryClass = {astro-ph.GA},
       adsurl = {https://ui.adsabs.harvard.edu/abs/2022ApJ...938..109F}
}

@ARTICLE{Forrest2020b,
       author = {{Forrest}, Ben and {Marsan}, Z. Cemile and {Annunziatella}, Marianna and {Wilson}, Gillian and {Muzzin}, Adam and {Marchesini}, Danilo and {Cooper}, M.~C. and {Chan}, Jeffrey C.~C. and {McConachie}, Ian and {Gomez}, Percy and {Kado-Fong}, Erin and {La Barbera}, Francesco and {Lange-Vagle}, Daniel and {Nantais}, Julie and {Nonino}, Mario and {Saracco}, Paolo and {Stefanon}, Mauro and {van der Burg}, Remco F.~J.},
        title = "{The Massive Ancient Galaxies at z > 3 NEar-infrared (MAGAZ3NE) Survey: Confirmation of Extremely Rapid Star Formation and Quenching Timescales for Massive Galaxies in the Early Universe}",
      journal = {\apj},
         year = 2020,
        month = nov,
       volume = {903},
       number = {1},
          eid = {47},
        pages = {47},
          doi = {10.3847/1538-4357/abb819},
archivePrefix = {arXiv},
       eprint = {2009.07281},
 primaryClass = {astro-ph.GA},
       adsurl = {https://ui.adsabs.harvard.edu/abs/2020ApJ...903...47F}
}

@ARTICLE{McCracken2012,
       author = {{McCracken}, H.~J. and {Milvang-Jensen}, B. and {Dunlop}, J. and {Franx}, M. and {Fynbo}, J.~P.~U. and {Le F{\`e}vre}, O. and {Holt}, J. and {Caputi}, K.~I. and {Goranova}, Y. and {Buitrago}, F. and {Emerson}, J.~P. and {Freudling}, W. and {Hudelot}, P. and {L{\'o}pez-Sanjuan}, C. and {Magnard}, F. and {Mellier}, Y. and {M{\o}ller}, P. and {Nilsson}, K.~K. and {Sutherland}, W. and {Tasca}, L. and {Zabl}, J.},
        title = "{UltraVISTA: a new ultra-deep near-infrared survey in COSMOS}",
      journal = {\aap},
         year = 2012,
        month = aug,
       volume = {544},
          eid = {A156},
        pages = {A156},
          doi = {10.1051/0004-6361/201219507},
archivePrefix = {arXiv},
       eprint = {1204.6586},
 primaryClass = {astro-ph.CO},
       adsurl = {https://ui.adsabs.harvard.edu/abs/2012A&A...544A.156M}
}

@ARTICLE{Schinnerer2010,
       author = {{Schinnerer}, E. and {Sargent}, M.~T. and {Bondi}, M. and {Smol{\v{c}}i{\'c}}, V. and {Datta}, A. and {Carilli}, C.~L. and {Bertoldi}, F. and {Blain}, A. and {Ciliegi}, P. and {Koekemoer}, A. and {Scoville}, N.~Z.},
        title = "{The VLA-COSMOS Survey. IV. Deep Data and Joint Catalog}",
      journal = {\apjs},
         year = 2010,
        month = jun,
       volume = {188},
       number = {2},
        pages = {384-404},
          doi = {10.1088/0067-0049/188/2/384},
archivePrefix = {arXiv},
       eprint = {1005.1641},
 primaryClass = {astro-ph.CO},
       adsurl = {https://ui.adsabs.harvard.edu/abs/2010ApJS..188..384S}
}

@ARTICLE{LeFloch2009,
       author = {{Le Floc'h}, Emeric and {Aussel}, Herv{\'e} and {Ilbert}, Olivier and {Riguccini}, Laurie and {Frayer}, David T. and {Salvato}, Mara and {Arnouts}, Stephane and {Surace}, Jason and {Feruglio}, Chiara and {Rodighiero}, Giulia and {Capak}, Peter and {Kartaltepe}, Jeyhan and {Heinis}, Sebastien and {Sheth}, Kartik and {Yan}, Lin and {McCracken}, Henry Joy and {Thompson}, David and {Sanders}, David and {Scoville}, Nick and {Koekemoer}, Anton},
        title = "{Deep Spitzer 24 {\ensuremath{\mu}}m COSMOS Imaging. I. The Evolution of Luminous Dusty Galaxies{\textemdash}Confronting the Models}",
      journal = {\apj},
         year = 2009,
        month = sep,
       volume = {703},
       number = {1},
        pages = {222-239},
          doi = {10.1088/0004-637X/703/1/222},
archivePrefix = {arXiv},
       eprint = {0909.4303},
 primaryClass = {astro-ph.CO},
       adsurl = {https://ui.adsabs.harvard.edu/abs/2009ApJ...703..222L}
}

@ARTICLE{Smolvcic2017,
       author = {{Smol{\v{c}}i{\'c}}, V. and {Novak}, M. and {Bondi}, M. and {Ciliegi}, P. and {Mooley}, K.~P. and {Schinnerer}, E. and {Zamorani}, G. and {Navarrete}, F. and {Bourke}, S. and {Karim}, A. and {Vardoulaki}, E. and {Leslie}, S. and {Delhaize}, J. and {Carilli}, C.~L. and {Myers}, S.~T. and {Baran}, N. and {Delvecchio}, I. and {Miettinen}, O. and {Banfield}, J. and {Balokovi{\'c}}, M. and {Bertoldi}, F. and {Capak}, P. and {Frail}, D.~A. and {Hallinan}, G. and {Hao}, H. and {Herrera Ruiz}, N. and {Horesh}, A. and {Ilbert}, O. and {Intema}, H. and {Jeli{\'c}}, V. and {Kl{\"o}ckner}, H. -R. and {Krpan}, J. and {Kulkarni}, S.~R. and {McCracken}, H. and {Laigle}, C. and {Middleberg}, E. and {Murphy}, E.~J. and {Sargent}, M. and {Scoville}, N.~Z. and {Sheth}, K.},
        title = "{The VLA-COSMOS 3 GHz Large Project: Continuum data and source catalog release}",
      journal = {\aap},
         year = 2017,
        month = jun,
       volume = {602},
          eid = {A1},
        pages = {A1},
          doi = {10.1051/0004-6361/201628704},
archivePrefix = {arXiv},
       eprint = {1703.09713},
 primaryClass = {astro-ph.GA},
       adsurl = {https://ui.adsabs.harvard.edu/abs/2017A&A...602A...1S}
}

@ARTICLE{Chabrier2003,
       author = {{Chabrier}, Gilles},
        title = "{Galactic Stellar and Substellar Initial Mass Function}",
      journal = {\pasp},
         year = 2003,
        month = jul,
       volume = {115},
       number = {809},
        pages = {763-795},
          doi = {10.1086/376392},
archivePrefix = {arXiv},
       eprint = {astro-ph/0304382},
 primaryClass = {astro-ph},
       adsurl = {https://ui.adsabs.harvard.edu/abs/2003PASP..115..763C}
}

@ARTICLE{Stalevski2016,
       author = {{Stalevski}, Marko and {Ricci}, Claudio and {Ueda}, Yoshihiro and {Lira}, Paulina and {Fritz}, Jacopo and {Baes}, Maarten},
        title = "{The dust covering factor in active galactic nuclei}",
      journal = {\mnras},
         year = 2016,
        month = may,
       volume = {458},
       number = {3},
        pages = {2288-2302},
          doi = {10.1093/mnras/stw444},
archivePrefix = {arXiv},
       eprint = {1602.06954},
 primaryClass = {astro-ph.GA},
       adsurl = {https://ui.adsabs.harvard.edu/abs/2016MNRAS.458.2288S}
}

@ARTICLE{Stalevski2012,
       author = {{Stalevski}, Marko and {Fritz}, Jacopo and {Baes}, Maarten and {Nakos}, Theodoros and {Popovi{\'c}}, Luka {\v{C}}.},
        title = "{3D radiative transfer modelling of the dusty tori around active galactic nuclei as a clumpy two-phase medium}",
      journal = {\mnras},
         year = 2012,
        month = mar,
       volume = {420},
       number = {4},
        pages = {2756-2772},
          doi = {10.1111/j.1365-2966.2011.19775.x},
archivePrefix = {arXiv},
       eprint = {1109.1286},
 primaryClass = {astro-ph.CO},
       adsurl = {https://ui.adsabs.harvard.edu/abs/2012MNRAS.420.2756S}
}

@ARTICLE{Draine2014,
       author = {{Draine}, B.~T. and {Aniano}, G. and {Krause}, Oliver and {Groves}, Brent and {Sandstrom}, Karin and {Braun}, Robert and {Leroy}, Adam and {Klaas}, Ulrich and {Linz}, Hendrik and {Rix}, Hans-Walter and {Schinnerer}, Eva and {Schmiedeke}, Anika and {Walter}, Fabian},
        title = "{Andromeda's Dust}",
      journal = {\apj},
         year = 2014,
        month = jan,
       volume = {780},
       number = {2},
          eid = {172},
        pages = {172},
          doi = {10.1088/0004-637X/780/2/172},
archivePrefix = {arXiv},
       eprint = {1306.2304},
 primaryClass = {astro-ph.CO},
       adsurl = {https://ui.adsabs.harvard.edu/abs/2014ApJ...780..172D}
}

@ARTICLE{Calzetti2000,
       author = {{Calzetti}, Daniela and {Armus}, Lee and {Bohlin}, Ralph C. and {Kinney}, Anne L. and {Koornneef}, Jan and {Storchi-Bergmann}, Thaisa},
        title = "{The Dust Content and Opacity of Actively Star-forming Galaxies}",
      journal = {\apj},
         year = 2000,
        month = apr,
       volume = {533},
       number = {2},
        pages = {682-695},
          doi = {10.1086/308692},
archivePrefix = {arXiv},
       eprint = {astro-ph/9911459},
 primaryClass = {astro-ph},
       adsurl = {https://ui.adsabs.harvard.edu/abs/2000ApJ...533..682C}
}

@ARTICLE{BruzualCharlot03,
       author = {{Bruzual}, G. and {Charlot}, S.},
        title = "{Stellar population synthesis at the resolution of 2003}",
      journal = {\mnras},
         year = 2003,
        month = oct,
       volume = {344},
       number = {4},
        pages = {1000-1028},
          doi = {10.1046/j.1365-8711.2003.06897.x},
archivePrefix = {arXiv},
       eprint = {astro-ph/0309134},
 primaryClass = {astro-ph},
       adsurl = {https://ui.adsabs.harvard.edu/abs/2003MNRAS.344.1000B}
}

@ARTICLE{Yang2022,
       author = {{Yang}, Guang and {Boquien}, M{\'e}d{\'e}ric and {Brandt}, W.~N. and {Buat}, V{\'e}ronique and {Burgarella}, Denis and {Ciesla}, Laure and {Lehmer}, Bret D. and {Ma{\l}ek}, Katarzyna and {Mountrichas}, George and {Papovich}, Casey and {Pons}, Estelle and {Stalevski}, Marko and {Theul{\'e}}, Patrice and {Zhu}, Shifu},
        title = "{Fitting AGN/Galaxy X-Ray-to-radio SEDs with CIGALE and Improvement of the Code}",
      journal = {\apj},
         year = 2022,
        month = mar,
       volume = {927},
       number = {2},
          eid = {192},
        pages = {192},
          doi = {10.3847/1538-4357/ac4971},
archivePrefix = {arXiv},
       eprint = {2201.03718},
 primaryClass = {astro-ph.GA},
       adsurl = {https://ui.adsabs.harvard.edu/abs/2022ApJ...927..192Y}
}

@ARTICLE{Jin2018,
       author = {{Jin}, Shuowen and {Daddi}, Emanuele and {Liu}, Daizhong and {Smol{\v{c}}i{\'c}}, Vernesa and {Schinnerer}, Eva and {Calabr{\`o}}, Antonello and {Gu}, Qiusheng and {Delhaize}, Jacinta and {Delvecchio}, Ivan and {Gao}, Yu and {Salvato}, Mara and {Puglisi}, Annagrazia and {Dickinson}, Mark and {Bertoldi}, Frank and {Sargent}, Mark and {Novak}, Mladen and {Magdis}, Georgios and {Aretxaga}, Itziar and {Wilson}, Grant W. and {Capak}, Peter},
        title = "{{\textquotedblleft}Super-deblended{\textquotedblright} Dust Emission in Galaxies. II. Far-IR to (Sub)millimeter Photometry and High-redshift Galaxy Candidates in the Full COSMOS Field}",
      journal = {\apj},
         year = 2018,
        month = sep,
       volume = {864},
       number = {1},
          eid = {56},
        pages = {56},
          doi = {10.3847/1538-4357/aad4af},
archivePrefix = {arXiv},
       eprint = {1807.04697},
 primaryClass = {astro-ph.GA},
       adsurl = {https://ui.adsabs.harvard.edu/abs/2018ApJ...864...56J}
}

@ARTICLE{Marsan2022,
       author = {{Marsan}, Z. Cemile and {Muzzin}, Adam and {Marchesini}, Danilo and {Stefanon}, Mauro and {Martis}, Nicholas and {Annunziatella}, Marianna and {Chan}, Jeffrey C.~C. and {Cooper}, Michael C. and {Forrest}, Ben and {Gomez}, Percy and {McConachie}, Ian and {Wilson}, Gillian},
        title = "{The Number Densities and Stellar Populations of Massive Galaxies at 3 < z < 6: A Diverse, Rapidly Forming Population in the Early Universe}",
      journal = {\apj},
         year = 2022,
        month = jan,
       volume = {924},
       number = {1},
          eid = {25},
        pages = {25},
          doi = {10.3847/1538-4357/ac312a},
archivePrefix = {arXiv},
       eprint = {2010.04725},
 primaryClass = {astro-ph.GA},
       adsurl = {https://ui.adsabs.harvard.edu/abs/2022ApJ...924...25M}
}

@ARTICLE{Muzzin2013a,
       author = {{Muzzin}, Adam and {Marchesini}, Danilo and {Stefanon}, Mauro and {Franx}, Marijn and {Milvang-Jensen}, Bo and {Dunlop}, James S. and {Fynbo}, J.~P.~U. and {Brammer}, Gabriel and {Labb{\'e}}, Ivo and {van Dokkum}, Pieter},
        title = "{A Public K$_{s}$ -selected Catalog in the COSMOS/ULTRAVISTA Field: Photometry, Photometric Redshifts, and Stellar Population Parameters}",
      journal = {\apjs},
         year = 2013,
        month = may,
       volume = {206},
       number = {1},
          eid = {8},
        pages = {8},
          doi = {10.1088/0067-0049/206/1/8},
archivePrefix = {arXiv},
       eprint = {1303.4410},
 primaryClass = {astro-ph.CO},
       adsurl = {https://ui.adsabs.harvard.edu/abs/2013ApJS..206....8M}
}

@ARTICLE{Boquien2019,
       author = {{Boquien}, M. and {Burgarella}, D. and {Roehlly}, Y. and {Buat}, V. and {Ciesla}, L. and {Corre}, D. and {Inoue}, A.~K. and {Salas}, H.},
        title = "{CIGALE: a python Code Investigating GALaxy Emission}",
      journal = {\aap},
         year = 2019,
        month = feb,
       volume = {622},
          eid = {A103},
        pages = {A103},
          doi = {10.1051/0004-6361/201834156},
archivePrefix = {arXiv},
       eprint = {1811.03094},
 primaryClass = {astro-ph.GA},
       adsurl = {https://ui.adsabs.harvard.edu/abs/2019A&A...622A.103B}
}

@ARTICLE{Forrest2020a,
       author = {{Forrest}, Ben and {Annunziatella}, Marianna and {Wilson}, Gillian and {Marchesini}, Danilo and {Muzzin}, Adam and {Cooper}, M.~C. and {Marsan}, Z. Cemile and {McConachie}, Ian and {Chan}, Jeffrey C.~C. and {Gomez}, Percy and {Kado-Fong}, Erin and {L Barbera}, Francesco and {Labb{\'e}}, Ivo and {Lange-Vagle}, Daniel and {Nantais}, Julie and {Nonino}, Mario and {Pe{\~n}a}, Theodore and {Saracco}, Paolo and {Stefanon}, Mauro and {van der Burg}, Remco F.~J.},
        title = "{An Extremely Massive Quiescent Galaxy at z = 3.493: Evidence of Insufficiently Rapid Quenching Mechanisms in Theoretical Models}",
      journal = {\apjl},
         year = 2020,
        month = feb,
       volume = {890},
       number = {1},
          eid = {L1},
        pages = {L1},
          doi = {10.3847/2041-8213/ab5b9f},
archivePrefix = {arXiv},
       eprint = {1910.10158},
 primaryClass = {astro-ph.GA},
       adsurl = {https://ui.adsabs.harvard.edu/abs/2020ApJ...890L...1F}
}

@ARTICLE{2013A&A...558A..33A,
       author = {{Astropy Collaboration} and {Robitaille}, Thomas P. and
         {Tollerud}, Erik J. and {Greenfield}, Perry and {Droettboom}, Michael and
         {Bray}, Erik and {Aldcroft}, Tom and {Davis}, Matt and
         {Ginsburg}, Adam and {Price-Whelan}, Adrian M. and
         {Kerzendorf}, Wolfgang E. and {Conley}, Alexander and {Crighton}, Neil and
         {Barbary}, Kyle and {Muna}, Demitri and {Ferguson}, Henry and
         {Grollier}, Fr{\'e}d{\'e}ric and {Parikh}, Madhura M. and
         {Nair}, Prasanth H. and {Unther}, Hans M. and {Deil}, Christoph and
         {Woillez}, Julien and {Conseil}, Simon and {Kramer}, Roban and
         {Turner}, James E.~H. and {Singer}, Leo and {Fox}, Ryan and
         {Weaver}, Benjamin A. and {Zabalza}, Victor and {Edwards}, Zachary I. and
         {Azalee Bostroem}, K. and {Burke}, D.~J. and {Casey}, Andrew R. and
         {Crawford}, Steven M. and {Dencheva}, Nadia and {Ely}, Justin and
         {Jenness}, Tim and {Labrie}, Kathleen and {Lim}, Pey Lian and
         {Pierfederici}, Francesco and {Pontzen}, Andrew and {Ptak}, Andy and
         {Refsdal}, Brian and {Servillat}, Mathieu and {Streicher}, Ole},
        title = "{Astropy: A community Python package for astronomy}",
      journal = {\aap},
         year = "2013",
        month = "Oct",
       volume = {558},
          eid = {A33},
        pages = {A33},
          doi = {10.1051/0004-6361/201322068},
archivePrefix = {arXiv},
       eprint = {1307.6212},
 primaryClass = {astro-ph.IM},
       adsurl = {https://ui.adsabs.harvard.edu/abs/2013A&A...558A..33A}
}

@ARTICLE{2022ApJ...935..167A,
       author = {{Astropy Collaboration} and {Price-Whelan}, Adrian M. and {Lim}, Pey Lian and {Earl}, Nicholas and {Starkman}, Nathaniel and {Bradley}, Larry and {Shupe}, David L. and {Patil}, Aarya A. and {Corrales}, Lia and {Brasseur}, C.~E. and {N{\"o}the}, Maximilian and {Donath}, Axel and {Tollerud}, Erik and {Morris}, Brett M. and {Ginsburg}, Adam and {Vaher}, Eero and {Weaver}, Benjamin A. and {Tocknell}, James and {Jamieson}, William and {van Kerkwijk}, Marten H. and {Robitaille}, Thomas P. and {Merry}, Bruce and {Bachetti}, Matteo and {G{\"u}nther}, H. Moritz and {Aldcroft}, Thomas L. and {Alvarado-Montes}, Jaime A. and {Archibald}, Anne M. and {B{\'o}di}, Attila and {Bapat}, Shreyas and {Barentsen}, Geert and {Baz{\'a}n}, Juanjo and {Biswas}, Manish and {Boquien}, M{\'e}d{\'e}ric and {Burke}, D.~J. and {Cara}, Daria and {Cara}, Mihai and {Conroy}, Kyle E. and {Conseil}, Simon and {Craig}, Matthew W. and {Cross}, Robert M. and {Cruz}, Kelle L. and {D'Eugenio}, Francesco and {Dencheva}, Nadia and {Devillepoix}, Hadrien A.~R. and {Dietrich}, J{\"o}rg P. and {Eigenbrot}, Arthur Davis and {Erben}, Thomas and {Ferreira}, Leonardo and {Foreman-Mackey}, Daniel and {Fox}, Ryan and {Freij}, Nabil and {Garg}, Suyog and {Geda}, Robel and {Glattly}, Lauren and {Gondhalekar}, Yash and {Gordon}, Karl D. and {Grant}, David and {Greenfield}, Perry and {Groener}, Austen M. and {Guest}, Steve and {Gurovich}, Sebastian and {Handberg}, Rasmus and {Hart}, Akeem and {Hatfield-Dodds}, Zac and {Homeier}, Derek and {Hosseinzadeh}, Griffin and {Jenness}, Tim and {Jones}, Craig K. and {Joseph}, Prajwel and {Kalmbach}, J. Bryce and {Karamehmetoglu}, Emir and {Ka{\l}uszy{\'n}ski}, Miko{\l}aj and {Kelley}, Michael S.~P. and {Kern}, Nicholas and {Kerzendorf}, Wolfgang E. and {Koch}, Eric W. and {Kulumani}, Shankar and {Lee}, Antony and {Ly}, Chun and {Ma}, Zhiyuan and {MacBride}, Conor and {Maljaars}, Jakob M. and {Muna}, Demitri and {Murphy}, N.~A. and {Norman}, Henrik and {O'Steen}, Richard and {Oman}, Kyle A. and {Pacifici}, Camilla and {Pascual}, Sergio and {Pascual-Granado}, J. and {Patil}, Rohit R. and {Perren}, Gabriel I. and {Pickering}, Timothy E. and {Rastogi}, Tanuj and {Roulston}, Benjamin R. and {Ryan}, Daniel F. and {Rykoff}, Eli S. and {Sabater}, Jose and {Sakurikar}, Parikshit and {Salgado}, Jes{\'u}s and {Sanghi}, Aniket and {Saunders}, Nicholas and {Savchenko}, Volodymyr and {Schwardt}, Ludwig and {Seifert-Eckert}, Michael and {Shih}, Albert Y. and {Jain}, Anany Shrey and {Shukla}, Gyanendra and {Sick}, Jonathan and {Simpson}, Chris and {Singanamalla}, Sudheesh and {Singer}, Leo P. and {Singhal}, Jaladh and {Sinha}, Manodeep and {Sip{\H{o}}cz}, Brigitta M. and {Spitler}, Lee R. and {Stansby}, David and {Streicher}, Ole and {{\v{S}}umak}, Jani and {Swinbank}, John D. and {Taranu}, Dan S. and {Tewary}, Nikita and {Tremblay}, Grant R. and {de Val-Borro}, Miguel and {Van Kooten}, Samuel J. and {Vasovi{\'c}}, Zlatan and {Verma}, Shresth and {de Miranda Cardoso}, Jos{\'e} Vin{\'\i}cius and {Williams}, Peter K.~G. and {Wilson}, Tom J. and {Winkel}, Benjamin and {Wood-Vasey}, W.~M. and {Xue}, Rui and {Yoachim}, Peter and {Zhang}, Chen and {Zonca}, Andrea and {Astropy Project Contributors}},
        title = "{The Astropy Project: Sustaining and Growing a Community-oriented Open-source Project and the Latest Major Release (v5.0) of the Core Package}",
      journal = {\apj},
         year = 2022,
        month = aug,
       volume = {935},
       number = {2},
          eid = {167},
        pages = {167},
          doi = {10.3847/1538-4357/ac7c74},
archivePrefix = {arXiv},
       eprint = {2206.14220},
 primaryClass = {astro-ph.IM},
       adsurl = {https://ui.adsabs.harvard.edu/abs/2022ApJ...935..167A}
}
\bibliographystyle{aasjournalv7}

\end{document}